\documentclass[final,5p,times,twocolumn]{elsarticle}
\usepackage{xcolor}
\usepackage{amsmath}
\usepackage{dcolumn}
\usepackage{siunitx}

\usepackage{graphicx}
\usepackage{amssymb}
\usepackage{amsthm}

\usepackage{multicol}
\usepackage[export]{adjustbox}

\journal{Physics Letters B}

\begin{document}

\begin{frontmatter}

\title{Precision measurement of the magnetic octupole moment in $^{45}$Sc as a test for state-of-the-art atomic- and nuclear-structure theory}

\author[1]{R. P. de Groote}{\ead{ruben.p.degroote@jyu.fi}}
\author[1]{J. Moreno}
\author[2,3]{J. Dobaczewski}
\author[4]{\'{A}. Koszor\'{u}s}
\author[1]{I. Moore}
\author[1]{M. Reponen}
\author[5]{B. K. Sahoo}
\author[6]{C. Yuan}

\address[1]{Department of Physics, University of Jyv\"askyl\"a, PB 35(YFL) FIN-40351 Jyv\"askyl\"a, Finland}
\address[2]{Department of Physics, University of York, Heslington, York YO10 5DD, United Kingdom}
\address[3]{Institute of Theoretical Physics, Faculty of Physics, University of Warsaw, ul. Pasteura 5, PL-02-093 Warsaw, Poland}
\address[4]{Department of Physics, University of Liverpool, Liverpool L69 7ZE, United Kingdom}
\address[5]{Atomic, Molecular and Optical Physics Division, Physical Research Laboratory, Navrangpura, Ahmedabad 380009, India}
\address[6]{Sino-French Institute of Nuclear Engineering and Technology, Sun Yat-Sen University, Zhuhai 519082, China}

\begin{abstract}
We report on measurements of the hyperfine $A, B$ and $C$-constants of the $3d4s^2 ~^2D_{5/2}$ and $3d4s^2 ~^2D_{3/2}$ atomic states in $^{45}$Sc. High-precision atomic calculations of the hyperfine fields of these states and second-order corrections are performed, and are used to extract $C_{5/2}=-0.06(6)$\,kHz and $C_{3/2}=+0.04(3)$\,kHz from the data. These results are one order of magnitude more precise than the available literature. From the combined analysis of both atomic states, we infer the nuclear magnetic octupole moment $\Omega = -0.07(53) \mu_N b$, including experimental and atomic structure-related uncertainties. With a single valence proton outside of a magic calcium core, scandium is ideally suited to test a variety of nuclear models, and to investigate in-depth the many intriguing nuclear structure phenomena observed within the neighboring isotopes of calcium. We perform nuclear shell-model calculations of $\Omega$, and furthermore explore the use of Density Functional Theory for evaluating $\Omega$. From this, mutually consistent theoretical values of $\Omega$ are obtained, which are in agreement with the experimental value. This confirms atomic structure calculations possess the accuracy and precision required for magnetic octupole moment measurements, and shows that modern nuclear theory is capable of providing meaningful insight into this largely unexplored observable. 
\end{abstract}


\end{frontmatter}

\section{Introduction}

The application of laser spectroscopic techniques to elucidate the subtle perturbations of atomic energy levels due to the nuclear electromagnetic properties has given rise to the study of fundamental nuclear structure, in particular magnetic dipole moments ($\mu$), electric quadrupole moments ($Q$) and changes in the mean-squared nuclear charge radii $\delta \left\langle r^2 \right\rangle$. These methods, in combination with modern radioactive ion beam (RIB) facilities, offer a powerful probe of changes in the structure of exotic nuclei. They provide information on nuclear shell evolution, nuclear shapes and sizes, and single-particle correlations~\cite{neyens2005,flanagan2009,ruiz2016,marsh2018,ichikawa2018,miller2019}. 
% Measurements of charge radii have most recently been combined with the dipole polarizability $\alpha_{D}$, to benchmark theoretical calculations which constrain the neutron radius and neutron skin thickness of $^{68}$Ni~\cite{Kaufmann2020}.
The majority of the experimental techniques in current use at RIB facilities provide measurements of hyperfine frequency splittings with a precision of the order of 1 MHz~\cite{Campbell2016}. This limitation restricts the sensitivity to higher order terms in the electromagnetic multipole expansion of the nuclear current densities, as well as to higher order radial moments of the charge density distribution. The progress in the development of higher precision methods along with ongoing development of theoretical tools has the potential to provide new perspectives which could help shape our understanding of the atomic nucleus. 

\begin{figure*}[t]
    \centering
    \includegraphics[width=\textwidth]{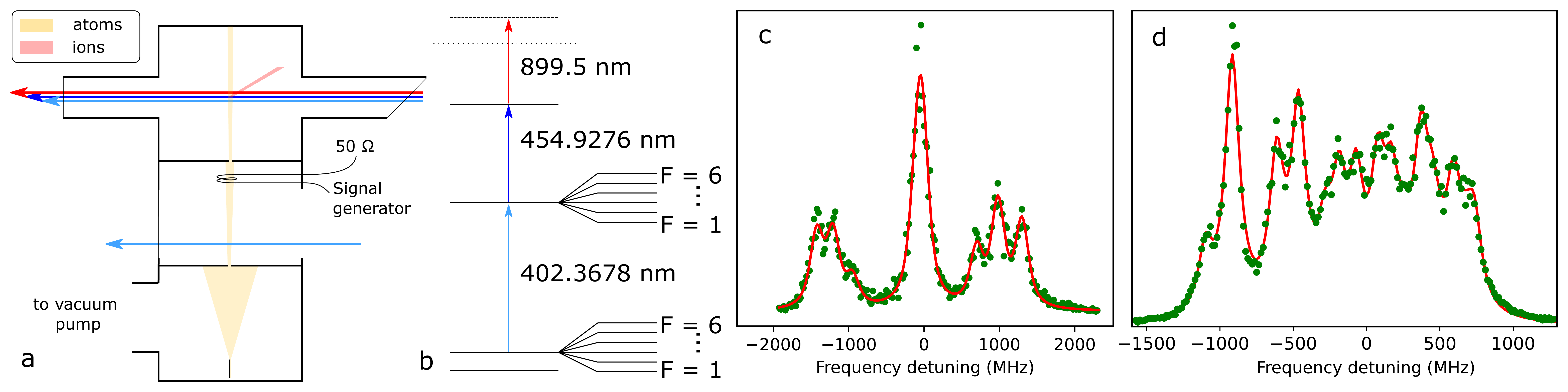}
    \caption{a) Schematic illustration of the experimental setup. The atom beam is produced from a tantalum oven mounted in the bottom vacuum vessel, intersects with the optical pumping laser beam, and then crosses the RF interaction region in the second vessel. After passing through a collimating slit, the atoms are then ionized using three lasers and subsequently counted using an ion detector. The laser ionization scheme used to study the D$_{5/2}$ state is shown in b), indicating also the hyperfine structure schematically. Figures c) and d) show an example spectra obtained by scanning the first step in the laser ionization scheme, for the transitions starting from the D$_{3/2}$ and D$_{5/2}$ states respectively.}
    \label{fig:expt}
\end{figure*}

Recently, high-precision isotope shift measurements combined with improved atomic calculations were proposed for a determination of the fourth-order radial moment of the charge density~\cite{Papoulia2016}, which can in turn be directly linked to the surface thickness of nuclear density~\cite{reinhard2020}. The hyperfine anomaly, only measured for a handful of radioactive isotopes (see e.g.~\cite{Takamine2014, papuga2014, zhang2015, schmidt2018, Persson2020}), would shed light on the distribution of magnetisation inside the nuclear volume~\cite{stroke2000,karpeshim2015}. In addition to the M1 and E2 moments, $\mu$ and $Q$ respectively, the M3 magnetic octupole moment $\Omega$ is in principle accessible using existing techniques for radioactive isotopes. To our knowledge, this observable has only been measured for 18 stable isotopes~\cite{Childs1971,Daly1954,Brown1966,Faust1963,Kusch1957,Jaccarino1954,Faust1961,Gerginov2003,Lewty2012,Unsworth1969,Singh2013,McDermott1960,Landman1970,fuller1976}. The general features of these values can be understood in terms of the Schwartz limits~\cite{Schwartz1955}. There are a few notable exceptions: the recently measured $\Omega$ of $^{133}$Cs~\cite{Gerginov2003} and $^{173}$Yb~\cite{Singh2013} are significantly larger than expected from shell model theory. For $^{173}$Yb, we recently performed an experiment to validate the earlier measurements, where a value of $\Omega$ which is zero within experimental uncertainties was obtained \cite{degroote2021}. In this Letter, we aim to further contribute to this ongoing work with an experimental and theoretical investigation of the hyperfine structure and nuclear electromagnetic moments of $^{45}$Sc. 

Our approach is threefold. Firstly, we describe a measurement protocol which combines the efficiency of resonance laser ionization spectroscopy (RIS) \cite{marsh2018,degroote2019,reponen2021} with the precision of radio-frequency (RF) spectroscopy \cite{Childs1992}. The efficiency provided by the RIS method is vital for future applications on radioactive isotopes due to limited production rates of radioactive ion beams at on-line facilities. The combination with RF spectroscopy offers a dramatic improvement in the precision as compared to conventional optical methods, by at least three orders of magnitude. We demonstrate this with a high-precision measurement of three nuclear electromagnetic moments of $^{45}$Sc, including $\Omega$.

Secondly, we combine these measurements with state-of-the-art atomic-structure calculations to evaluate the sensitivity of the $3d4s^2 \ ^2D_{3/2, 5/2}$ states in neutral scandium to the nuclear octupole moment $\Omega$. We evaluate the impact of off-diagonal HFS effects, essential to extract $\Omega$ from the measurements. We note that both the $D_{3/2}$ and meta-stable $D_{5/2}$ state are expected to be well-populated in a fast-beam charge exchange reaction~\cite{Vernon2019}. Therefore, radioactive scandium isotopes could be studied using collinear laser-double resonance methods~\cite{Nielsen1983, Childs1992} in the future.

Thirdly, with a single proton outside a doubly-magic calcium ($Z=20$) core, comparison of $\Omega$ for a chain of scandium isotopes provides a first important testing ground for nuclear theory calculations. Furthermore, such measurements could help shed light on the many intriguing nuclear structure phenomena observed in the calcium isotopes~\cite{steppenbeck2013,wienholtz2013,ruiz2016,Tanaka2020}. The proximity to proton- and neutron shell closures makes it possible to perform both e.g. shell-model and Density Functional Theory (DFT) calculations. As we seek to eventually examine all existing values of $\Omega$ in one consistent framework, with measurements for nuclei scattered throughout the nuclear landscape, developing a reliable global theory for magnetic properties would be highly advantageous. So far, very little is known regarding the overall performance of standard nuclear DFT in describing $\mu$, cf.~Refs.~\cite{(Ach14),(Bon15),(Bor17)}, and nothing is known about the DFT values of $\Omega$. Here, we thus start this investigation with $^{45}$Sc. The comparison to nuclear shell-model calculations, which have a more well-established track record in computing both $\mu$ and $\Omega$ (see e.g.~\cite{brown1980}), serves to benchmark these developments.

These three aspects are all required ingredients for a systematic study of $\Omega$ throughout the nuclear chart. The extraction of a higher-order electromagnetic moment from the evaluation of atomic spectra in a nuclear-model independent manner has the potential to provide new insight into the distribution of protons and neutrons within the nuclear volume. $\Omega$ is affected by correlations (core polarization and higher order configuration mixing) differently than the magnetic dipole moment, as was highlighted via calculations of the nuclear magnetization distribution of $^{209}$Bi~\cite{senkov2002}. Measurements of $\Omega$ may thus furthermore help to address open questions related to e.g. effective nucleon $g$-factors and charges.

\section{Overview of the experiment}

The value of $\Omega$ can be extracted from the first-order shift ($E_F^{(1)}$) in the hyperfine structure (HFS) interval, governed by the hyperfine interaction Hamiltonian:
\begin{align}
    \mathcal{H}_{\text{hyp}} & = A{\bf I \cdot J} + B\frac{{3({\bf I\cdot J})^2 + \frac {3}{2}({\bf I \cdot J})- I(I+1)J(J+1)}}{2I(2I-1)J(2J-1)} \notag \\ 
         + & C \left[ \frac{10({\bf I \cdot J})^3 + 20({\bf I \cdot J})^2 }{I(I-1)(2I-1)J(J-1)(2J-1)} \right.  \notag \\ 
         + &\left.\frac{2{\bf I \cdot J}\{ I(I+1) + J(J+1) - 3N + 3\} - 5N} {I(I-1)(2I-1)J(J-1)(2J-1)}\right],    \label{eq:hfs}
\end{align}
where $N = I(I+1)J(J+1)$, and noting $\left\langle F, m_F \right|  {\bf I \cdot J} \left| F, m_F\right\rangle = \frac12 [F(F+1) - I(I+1) - J(J + 1)]$. In these expressions, $I$, $J$ and $F$ are the nuclear, atomic, and total angular momentum, while $A$, $B$ and $C$ are the magnetic dipole (M1), electric quadrupole (E2) and magnetic octupole (M3) HFS constants, respectively. These are all proportional to their corresponding nuclear moment, in a way which depends on the field distribution generated by the electrons at the site of the nucleus. Thus, accurate atomic structure calculations of $C/ \Omega$ have to be performed to extract $\Omega$ from $C$. 

\begin{figure*}[t]
    \centering
    \includegraphics[width=0.9\textwidth]{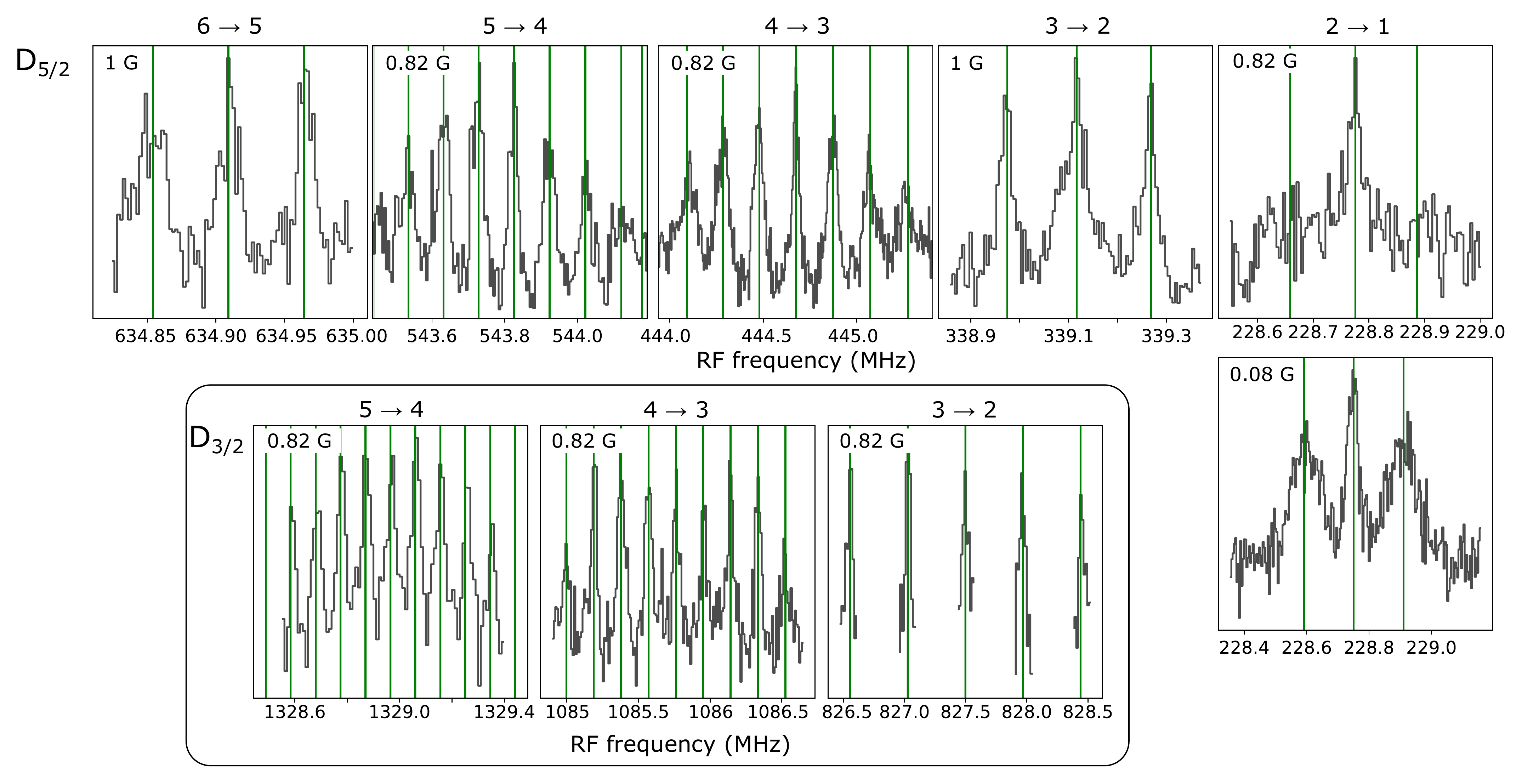}
    \caption{RF scans of several $(F, m_{F}) \rightarrow (F-1,m_{F})$ transitions in the D$_{5/2}$ and D$_{3/2}$ hyperfine manifolds. Green vertical lines indicate the $\Delta m_F = 0$ resonance locations governed by Eq.\eqref{eq:fullH} with the best-fitting hyperfine constants and magnetic field values. The y-axis represents the ratio of RF-on and RF-off datapoints, as described in the text.}
    \label{fig:rf_data}
\end{figure*}

% Laser-RF double-resonance methods have a well-proven track record in precise measurements of hyperfine structure constants~\cite{Childs1992}.
There are three stages in our experiment, schematically illustrated in Fig.~\ref{fig:expt}. First, by tuning a continuous wave (cw) laser into resonance with a transition from one of the hyperfine levels ($F$) of the atomic ground state into a corresponding hyperfine level of an excited $J$ state, population may be optically pumped. Through de-excitation from the excited state into either another level ($F'$) of the ground-state hyperfine manifold, or into other dark states, the population of the state $F$ is depleted. If RIS is subsequently performed starting from the same $F$ state, a reduced ion count rate is observed. If now, prior to the laser ionization stage, an RF field is tuned into resonance with a $(F, m_F) \rightarrow (F-1,m_{F})$ transition, the observed ion count rate increases. By scanning the frequency of the RF and recording the ion count rate, the hyperfine spacing between the levels $F$ and $F'$ of the ground-state manifold can thus be measured precisely. Due to the relative orientation of the oscillating magnetic field and the earth's magnetic field,  both pointing along the atom beam axis, only $\Delta m_F=0$ resonances are observed.

The vacuum chamber used for the experiments is shown in Fig.~\ref{fig:expt}a. It consists of three cylindrical vessels, one to produce the atom beam, a second one for optical pumping and RF spectroscopy, and the third and final one for laser ionization and ion detection. These vessels are separated by metal walls with a thin slit (1x20mm) used to collimate the atom beam. We produced an atomic beam of stable scandium in the bottom chamber by resistively heating a tantalum furnace. In the second chamber, up to 15\,mW of cw laser light crossed the atom beam orthogonally, in order to optically pump the atoms. This light was produced with a frequency-doubled Sirah Matisse Ti:Sapphire laser, focused to a $\sim$1\,mm spot. The laser was tuned to drive either the 25 014.190\,cm$^{-1}$ $3d4s({}^3D)4p \ ^2D^{\circ}_{5/2}$ state or the 24866.172\,cm$^{-1}$ $3d4s({}^3D)4p \ ^2D^{\circ}_{3/2}$, starting from respectively the thermally populated $3d4s^2 \ ^2D_{5/2}$ state at 168.3371\,cm$^{-1}$ and the $3d4s^2 \ ^2D_{3/2}$ ground state. 
The atomic beam then passed through a loop of wire, 8\,cm above the optical pumping region. This wire was terminated with 50 $\Omega$ in order to ensure good impedance matching, minimizing reflected RF power. The voltage standing wave ratio (VSWR) was measured using a Rhode\&Schwarz ZVL Network analyser, and was found to vary negligibly within the scan range. The generator is a DS instruments DS6000 pro PureSine signal generator, referenced to an internal 10\,MHz reference with a quoted accuracy of 280 parts per billion. For the measurements, the generator was set to output 5 mW of RF power. The atoms are exposed to the RF field for an estimated few 10\,$\mu$s, which thus leads to expected linewidths of a few 10\,kHz. 

The atoms are then further collimated and orthogonally overlapped with the ionization lasers which are focused into a 1x1mm$^2$ spot, 13\,cm above the RF interaction region, in the third chamber. A three-step resonant laser ionization scheme was used to ionize the scandium atoms, derived from the scheme in~\cite{raeder2013source}, shown in Fig.~\ref{fig:expt}b. The first step is provided using a $\sim$5~$\%$ pick-off from the cw laser beam used for the optical pumping stage. The other two steps were produced by pulsed Ti:Sapphire lasers (10\,kHz repetition rate), tuned to the 25014.190\,cm$^{-1} \rightarrow$ 46989.493\,cm$^{-1}$ transition or the 24866.172\,cm$^{-1} \rightarrow$ 46914.540\,cm$^{-1}$ transition, and to a broad auto-ionizing state at $\sim$58104\,cm$^{-1}$ or 58037\,cm$^{-1}$. The laser powers used for the laser ionization were approximately 0.75 mW, 50 mW and 500 mW for the first, second and third steps, respectively. 

Prior to performing any double-resonance measurements, an estimate of the HFS constants can be obtained by scanning the frequency of the first laser step, as shown in Fig.\,\ref{fig:expt}c,d. During the RF scans, the laser wavelength was kept fixed to pumping wavelengths suitable for the different RF lines, and the RF field was introduced and scanned. Fig.~\ref{fig:rf_data} shows examples of the RF lines which were obtained. The Zeeman splitting observed in wider-range scans can be used to determine the magnetic field strength. As the measurements presented in this work were performed over a time scale of two years, different values of this field are obtained between the different datasets: 1.03\,G for the first set of measurements, 80\,mG for a second set, and 0.83\,G for the third. The measurements with a field of 80\,mG were performed only for the $(1,0) \rightarrow (2,0)$ transition of the D$_{5/2}$ state, where the external field was partially shielded with mu-metal foils. This was done in order to evaluate possible systematic errors, since this line is more sensitive to the B-field than the others (e.g. at 1\,G the $m_F = 0 \rightarrow m_{F'}=0$ shifts by as much as 39\,kHz). The data from all measurements was found to be consistent, indicating the measurement protocol is reliable and the magnetic field strengths can be accurately assessed from the Zeeman splitting.

\section{Analysis}

Extracting accurate HFS constants requires atomic structure calculations to estimate the second-order shift ($E_F^{(2)}$) due to M1-M1, M1-E2 and E2-E2 interactions. These calculations will be discussed first. 

\subsection{Calculation of hyperfine constants and second-order shifts}

The relativistic coupled-cluster (RCC) theory, known as the gold-standard of many-body theory~\cite{shavitt2009}, is used to evaluate $C/ \Omega$ and the matrix elements involving the second-order hyperfine interaction Hamiltonians. In this work, we expand on earlier calculations~\cite{sahoo2005} presenting $A/g_I$ (with $g_I= \mu/I$), $B/Q$ and $C/ \Omega$ with a larger set of orbitals, using up to $19s$, $19p$, $19d$, $18f$, $17g$, $16h$ and $15i$ orbitals in the singles- and doubles-excitation approximation in the RCC theory (RCCSD method). Due to limitations in computational resources, we correlate electrons up to $g$-symmetry orbitals in the singles-, doubles- and triples-excitation approximation in the RCC theory (RCCSDT method). We quote the differences in the results from the RCCSD and RCCSDT methods as `$+$Triples'. Contributions from the Breit and lower-order quantum electrodynamics (QED) interactions are determined using the RCCSD method, and added to the final results as `$+$Breit' and `$+$QED', respectively. Contributions due to the Bohr-Weisskopf (BW) effect are estimated in the RCCSD method considering a Fermi-charge distribution within the nucleus and corrections are quoted as `$+$BW'. We also extrapolated contributions from an infinite set of basis functions and present these as `Extrapolation'. 

\begin{table*}[tb]
    \centering
    \small
    \caption{Theoretical HFS constants of the $3d4s^2 \ ^2D_{3/2,5/2}$ state. The dominant off-diagonal reduced matrix elements $T_k = \langle 3/2 ||T_e^{(k)}|| 5/2 \rangle = -\langle 5/2 ||T_e^{(k)}|| 3/2 \rangle$ required for the estimation of the second-order corrections to the hyperfine intervals are provided in the last two rows.} \label{tab:theor}
    \begin{tabular}{r|ccc|ccccc|rl}
	   & Dirac-Fock & RMBPT(2)  & RCCSD   &  + Triples &  + QED  &  + Breit &  + BW & Extrapolation & Total & \\
	  \hline
	  D$_{3/2}$  &&&&&&&&&&\\
      $A/\mu$ & 49.520 & 54.008 & 56.172 & 0.656 & 0.013 & 0.153 & -0.002 & 0.055   &   57.0(6)& MHz$/\mu_N$  \\
      $B/Q$  & 107.037& 122.824& 126.343& -0.787& 0.000& -0.046& 0.000& 0.000& 125(2) & MHz$/$b \\
	  $C/ \Omega $ &  1.91 & -5.85 & -4.86 & -0.65 & -0.02 & -0.35 & ~0 & 0.09 & -5.8(3) & $10^{-2}$ kHz/($\mu_N$ b) \\
	  \hline
	  D$_{5/2}$ &&&&&&&&&&\\
      $A/\mu$ & 21.066   &   19.744   &   22.416   &   0.179   &   0.013   &   0.057   &   0.002     &   0.021   &   22.7(4)& MHz$/\mu_N$  \\
      $B/Q$   & 151.39   &   173.84   &   176.11   &   -1.30   &    0.05   &   0.09    &   $\sim 0$   &   -0.60   &   175(2)& MHz$/$b \\
	  $C/ \Omega $   & 0.78  &   2.33 &  -17.09  &  0.58 &$\sim 0$&  0.80 &$\sim 0$&  - 0.14  &  -15.9(2) & $10^{-2}$ kHz/($\mu_N$ b) \\	  
	  \hline
	  $T_1$ & 83.52  & 176.18  & 145.03  & 10.15  &  -0.1  &   0.61 &        &  0.1   & 156(6) & MHz/$\mu_N$   \\
	  $T_2$ & 311.27 & 355.29  & 376.17  &  -11.27&   0.19 &   0.48 &        &   0.02 & 366(7) & MHz/b   \\
    \end{tabular}
\end{table*}

The $A/\mu$, $B/Q$ and $C/ \Omega$ values of the $3d4s^2 \ ^2D_{3/2, 5/2}$ states are tabulated in Table~\ref{tab:theor}. To obtain $A$ and $B$, listed in Table~\ref{tab:hyperfine}, recommended literature values of the moments were used ($\mu = +4.75400(2)\,\mu_N$~\cite{stone2019} and $Q = -0.220(9)$\,b~\cite{stone2016}). Uncertainties are estimated from the neglected higher-level excitations of the RCC theory. The shift $E_F^{(2)}$ due to M1-M1, M1-E2 and E2-E2 interaction terms is given by \cite{sahoo2015}:
\begin{eqnarray}
E_F^{(2)} &=& E_F^{M1-M1} + E_F^{M1-E2} + E_F^{E2-E2} \nonumber \\ 
&=& \sum_{J'} \left | \left  \{
                               \begin{matrix}
                                  F & J & I \cr
                                1 & I & J' \cr
                                 \end{matrix}  \right \}  \right |^2 \eta  \nonumber \\
    &+& \sum_{J'} \left  \{
                    \begin{matrix}
                      F & J & I \cr
                      1 & I & J' \cr
                     \end{matrix}  \right \}  \left  \{
                    \begin{matrix}
                      F & J & I \cr
                      2 & I & J' \cr
                     \end{matrix}  \right \} \zeta \nonumber \\
     &+& \sum_{J'}  \left | \left  \{
                               \begin{matrix}
                                  F & J & I \cr
                                2 & I & J' \cr
                                 \end{matrix}  \right \}  \right |^2 \epsilon  \label{eq:seqorder}
\end{eqnarray}
where 
\begin{eqnarray}
\eta &=&\frac{(I+1)(2I+1)}{I} \mu^2 \frac{|\langle J'||T_e^{(1)}||J \rangle|^2}{E_J -E_{J'}} , \nonumber \\
\zeta &=&\frac{(I+1)(2I+1)}{I} \sqrt{\frac{2I+3}{2I-1}} \mu Q \frac{\langle J'||T_e^{(1)}||J \rangle \langle J'||T_e^{(2)}||J \rangle}{E_J -E_{J'}} \nonumber \\
&& \text{and} \nonumber \\
\epsilon &=&\frac{(I+1)(2I+1)(2I+3)}{I(2I-1)} Q^2 \frac{|\langle J'||T_e^{(2)}||J \rangle|^2}{E_J -E_{J'}} . \nonumber
\end{eqnarray}
In these expressions, ${\bf T}_e^{(k)}$ is the spherical tensor operator with rank ``$k$ ($>0$)" in the electronic coordinates. We quote numerical values for these second-order matrix elements in Table~\ref{tab:theor}. We only consider the dominant contributing matrix elements between the $3d4s^2$  $^2D_{5/2}$ state and the $3d4s^2$  $^2D_{3/2}$ state. Intermediate results from the zeroth-order calculation using the Dirac-Fock method and the second-order relativistic many-body perturbation theory (RMBPT(2) method) are presented to demonstrate the propagation of electron correlation effects from lower to all-order RCC methods. 

\subsection{Analysis of hyperfine resonances}

The data is processed and analysed as follows. For each value of the rf frequency, the number of ion counts is recorded for a time interval of typically one second, once with the output of the rf generator on, and once with the output of the generator off. By repeating this procedure for the desired range of frequencies, a spectrum is obtained by taking the ratio of the two measured counts. If needed, a rebinning of the data is performed in order to improve the signal-to-noise ratio. All spectra obtained in this way are then fitted by explicitly diagonalizing the following Hamiltonian:
\begin{align}
    \mathcal{H} = \mathcal{H}_{\text{hyp}} + B_0 \cdot(g_J \mu_B J_z + g \mu_N I_z),\label{eq:fullH}
\end{align}
with $\mathcal{H}_{\text{hyp}}$ given in Eq.\eqref{eq:hfs} and $B_0$ the external magnetic field, and then correcting these eigenvalues using the expressions for the second-order shift given in Eq. \eqref{eq:seqorder}. Resonance locations can then be calculated as differences of these eigenvalues, $\Delta F = \pm1, \Delta m_F = 0, \pm 1$. The best-fitting values of the hyperfine constants $A$, $B$ and $C$ are found by comparing the calculated resonance locations with those observed in the experimental data using least-squares minimization. Additional free parameters in the fit are the value of the magnetic field $H$ and the heights of the resonances predicted by the above procedure, all of which are allowed to vary from one spectrum to the next in order to obtain the best goodness-of-fit. Note that the influence of the nuclear $g$-factor on the total Zeeman splitting is negligible, but the effect was included explicitly for completeness.

The hyperfine constants of $^{45}$Sc, with and without use of second-order shifts, are given in Table~\ref{tab:hyperfine}, and the values extracted for the octupole  moment are shown in Fig.~\ref{fig:comparison}. The systematic uncertainty due to the atomic calculations is given in square brackets. This uncertainty was estimated by the change in hyperfine constants obtained by varying the values of $T_1$ and $T_2$ within the theoretical error bar. Our results for $A,B$ and $C$ agree well with literature~\cite{Childs1971}, and are at least an order of magnitude more precise. 

\section{Results and interpretation}

\begin{table*}[ht!]
    \centering
    \small
    \caption{Experimental and theoretical HFS constants and $\Omega$ values, without and with the second-order corrections. HFS constants beyond the octupole term were found to be zero within errors, and were thus not included in the fit.} \label{tab:hyperfine}
    \begin{tabular}{r|r|c|cc|cc}
        & & Theory & \multicolumn{2}{c|}{ Expt. Ref.~\cite{Childs1971}} & \multicolumn{2}{c}{Expt. this work}   \\
        & & This work &  Uncorrected  & Corrected  & Uncorrected & Corrected  \\
            \hline
            D$_{3/2}$& A [MHz]           & 271(3)   & 269.556(1)   &  269.558(1)    & 269.55817(5)   & 269.55844(7)[3]    \\
            & B [MHz]           & -27.5(5) &  -26.346(4)  &    -26.360(8)  &   -26.3531(9)  &  -26.3596(5)[5]    \\
            & C [kHz]           &          &  --          &   --           &    -0.010(22)  &  0.039(28)[2]      \\
            & $\Omega\,[\mu_N$b]&          & --           & --             &  0.17(38)      &  -0.68(49)[6]      \\
            \hline
            D$_{5/2}$ & A [MHz]           & 108(2)   & 109.032(1)   &   109.033(1)   &  109.03275(7)   & 109.03297(5)[3]   \\
            & B [MHz]           & -38.5(5) &  -37.387(12) &    -37.373(15) &   -37.3954(12)  &  -37.3745(8)[15]  \\
            & C [kHz]           &          &  1.7(10)     &    1.5(12)     &    0.31(8)      &  -0.062(59)[17]   \\
            & $\Omega\,[\mu_N$b]&          & -10.7(63)    &  -9.4(75)      &   -1.92(51)     &  0.39(37)[11]       
    \end{tabular}
\end{table*}

\begin{table}[ht!]
    \centering
    \small
    \caption{Experimental and theoretical values of $\Omega$. The experimental value obtained in this work is the dispersion-corrected weighted mean of the values for the two $D_J$ states, where the total (statistical + systematic) was used in the weighting and to compute the total uncertainty.} \label{tab:omega}
    \begin{tabular}{c|r|c}
        &  & $\Omega$ [$\mu_N$b] \\
        \hline
        Expt. & Literature \cite{Childs1971}    & -9.4(75)          \\
        & This work  & -0.07(53)         \\
        \hline
        Theory & Schwartz $g_s=1/0.6$         & 0.65 / 0.46       \\
        & SM $g_s=1/0.6$     & 0.45(4) / 0.32(4) \\
        &  DFT                         & 0.245(17)         \\
    \end{tabular}
\end{table}

Since scandium has a single proton outside of the magic shell of $Z=20$, a single-particle shell model estimate for $\Omega$~\cite{Schwartz1955} would be expected to be fairly good. We find $\Omega_{\text{sm}} = 0.46 \ \mu_N b$, using $\left\langle r^2 \right\rangle^{1/2} = 4.139$\,fm as the radius of the $f_{7/2}$ orbit (obtained from DFT calculations discussed later). This value is in good agreement with the experimental values.
% Note that for $^{133}$Cs a discrepancy between the simple shell model and the experimental data was observed: $\Omega_{\text{exp}} = 0.82(10) \ \mu_N b$~\cite{Gerginov2003}, while $\Omega_{\text{sm}} = 0.14 \ \mu_N b$ using $\left\langle r^2 \right\rangle^{1/2} \approx R = 4.8041(46)$\,fm~\cite{angeli2013}. The single-particle shell model estimate for the dipole moment $\mu$ of this isotope agrees rather well with experiment, making the disagreement for $\Omega$ difficult to explain. 
As a step towards a more complete understanding of $\Omega$ for $^{45}$Sc, and as a step towards understanding this observable in general, we examine it in more detail using more realistic nuclear models. 

\subsection{Nuclear shell-model}

Shell-model calculations were performed using different interactions in a $(sd) pf$-shell model space ~\cite{PhysRevC.65.061301, EurPhysJA.25.499, PhysRevLett.66.1134, POVES2001157}. The values of $\Omega$ are calculated by the nuclear shell model through the code KSHELL \cite{SHIMIZU2019372}. The expression of $\Omega$ is defined as
\begin{eqnarray*}
\Omega=-M_{3}&=& -\sqrt{\frac{4\pi}{7}}\left(
                                 \begin{array}{ccc}
                                   J & 3 & J \\
                                   -J & 0 & J \\
                                 \end{array}
                               \right)  \nonumber \\
              &&\times(g^{(l)}_{p}l_{p} + g^{(l)}_{n}l_{n} + g^{(s)}_{p}s_{p} + g^{(s)}_{n}s_{n}) \nonumber 
\end{eqnarray*}
where $l_{p(n)}$ and $s_{p(n)}$ are the proton (neutron) angular momentum and spin terms of nuclear matrix elements, respectively, and $g^{(l)}_{p(n)}$ and $g^{(s)}_{p(n)}$ are corresponding proton (neutron) $g$ factors. The structure of $^{45}$Sc is calculated using seven Hamiltonians, GXPF1~\cite{PhysRevC.65.061301}, GXPF1A~\cite{EurPhysJA.25.499}, KB3~\cite{PhysRevLett.66.1134}, and KB3G~\cite{POVES2001157} for the $pf$-shell model space, and SDPF-M~\cite{PhysRevC.60.054315}, SDPF-MU~\cite{PhysRevC.86.051301}, and SDPFUSI~\cite{PhysRevC.79.014310} for the $sdpf$-shell model space. 
% The Hamiltonian is constructed based on the monopole-based universal interaction V$_{\text{MU}}$ \cite{otsuka2010} and the spin-orbit force from M3Y~\cite{m3y1977}(V$_{\text{MU}}$+LS). The strengths of all central, tensor, and spin-orbit parts are taken to be the same as the original values except for one central part C10 with T,S=1,0. The strength of the C10 part is enhanced by $15\%$ for the proton-proton interaction and $5\%$ for the neutron-neutron interaction, which provides better descriptions of nuclei around $^{132}$Sn \cite{yuan2018NPR}. The same nuclear interaction is used to explain the recently observed isomeric states in $^{123,125}$Ag \cite{PhysRevLett.122.212502}.

The $\Omega$ for $^{45}$Sc 
% and $^{133}$Cs 
is dominated by the proton contribution, 
% In $^{133}$Cs, the angular momentum and spin contribution to $\Omega$ have the opposite sign and thus largely cancel, while for $^{45}$Sc 
with the angular momentum and spin contributions having the same sign. 
% For $^{45}$Sc, 
We obtain values in the range 0.41-0.49\,$\mu_N$b with free $g$-factors and 0.28-0.35\,$\mu_N$b with a spin-quenching factor of 0.6 for the different shell model calculations. The inclusion of cross-shell excitations from the $sd$-shell to the $pf$-shell enhances the correlation beyond the single $f_{7/2}$ proton configuration, which results in small increases in $\Omega$. 
% In contrast to $^{45}$Sc, the calculated $\Omega$ for $^{133}$Cs is very sensitive to the choice of effective $g$-factors: 0.02$\mu_N$b with free $g$-factors and 0.16$\mu_N$b with a spin-quenching factor of 0.6. While all shell-model results agree well with experiment for $^{45}$Sc, they underestimate the value of $\Omega$ for $^{133}$Cs. 

\subsection{Nuclear Density Functional Theory}

\begin{figure}[tb]
\begin{center}
\includegraphics[width=\columnwidth]{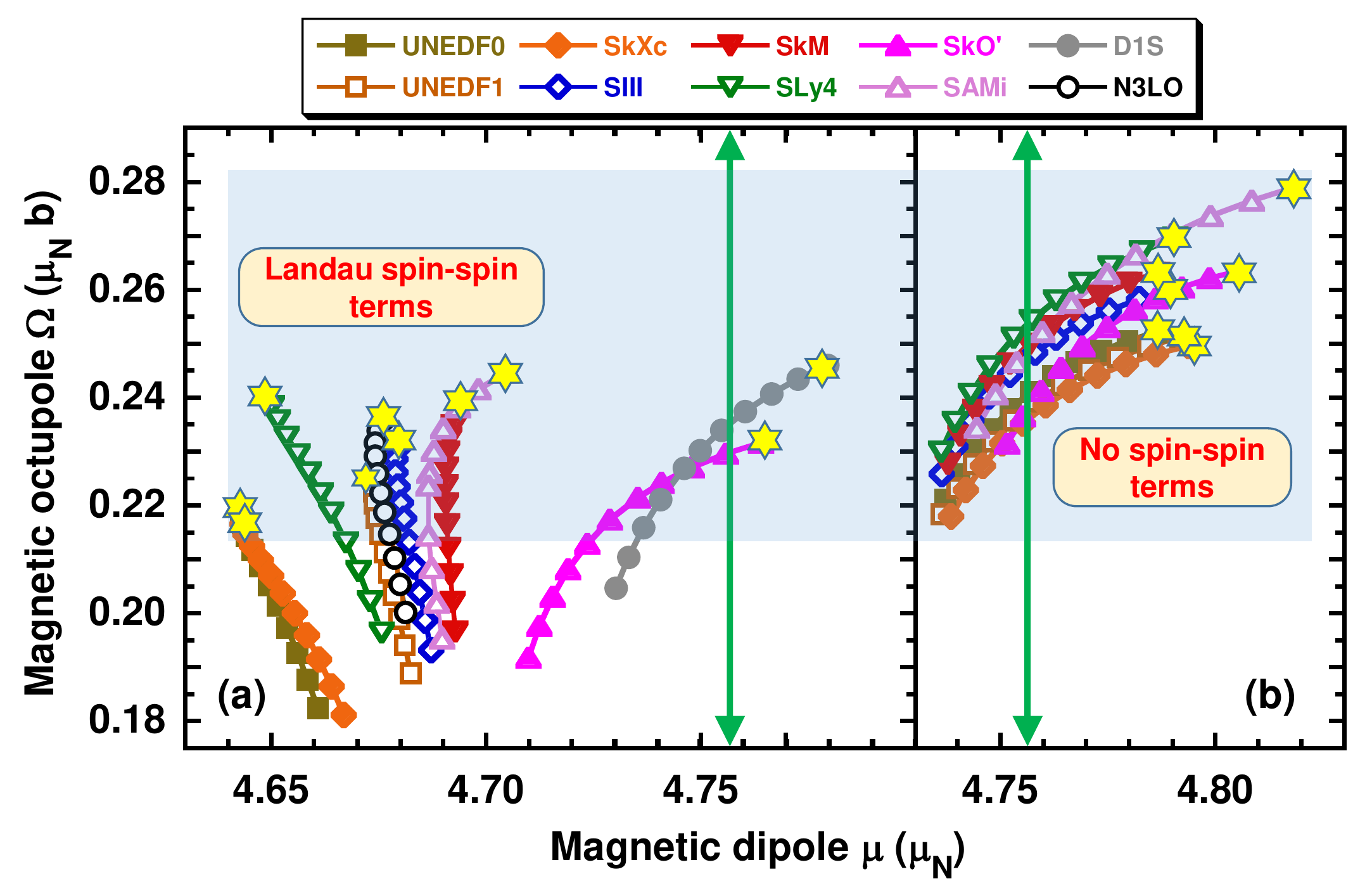}
\caption{Values of $\mu$ and $\Omega$ of the $I=7/2^-$ angular-momentum-projected ground states of $^{45}$Sc. Panels (a) and (b) show results obtained with Skyrme functionals supplemented by the Landau spin-spin terms and with no spin-spin terms, respectively. Arrows mark the
experimental value of $\mu$ and visualize the experimental error bars of $\Omega=-0.07(53)$, which are outside the scale of the figure.
\label{fig:M1-M3}
}
\end{center}
\end{figure}

% The discussion of $\Omega$ from a DFT perspective will be restricted to $^{45}$Sc, since the proximity of $^{45}$Sc to the semi-magic Ca simplifies the theory. 
We determined values of $\mu$, $Q$, and $\Omega$ for oblate states in $^{45}$Sc. We used constrained intrinsic mass quadrupole moments $Q_{20}=\langle2z^2-x^2-y^2\rangle$ varying between $-$1\,b and 0, with points at $-$1\,b marked by stars, see Fig.~\ref{fig:M1-M3}. The obtained unpaired mean-field solutions were projected on the $I=7/2^-$ ground-state angular momentum. Proton and neutron configurations were fixed at $\pi3^1$ and $\nu3^4$, where $3^n$ represents the occupied $n$ lowest oblate orbitals in the $\ell=3$ $f_{7/2}$ shell. No effective charges or effective  $g$-factors were used. 

Results of the DFT calculations were obtained
using the code {\sc hfodd} (version 2.95j) \cite{(Dob21f)}. To
represent single-particle wave functions, we used the basis of
$N_0=14$ spherical harmonic oscillator shells. We run the code in the
mode of conserved parity along with broken simplex and broken time reversal.
We used an infinitesimal angular frequency of $\hbar\omega=1$\,keV
aligned along the $z$ direction. Simultaneously, the nucleus was oriented
in space so that the axial-symmetry axis was also aligned along the
$z$ direction. This allowed for splitting single-particle energies
according to their projections of the angular momentum $K$ on the
symmetry axis, without affecting their wave functions. At the same
time, all single-particle wave functions acquired good $K$ quantum
numbers.

To stabilize the convergence, during the self-consistent
iterations the total wave functions were additionally projected on
the axial symmetry \cite{(Dob21f)}. Occupied single-particle wave functions were fixed by distributing the neutrons and protons according to the partitions of numbers of occupied states in
individual blocks of given $K$ \cite{(Dob21f)}. This defined specific
intrinsic configurations $\pi3^1$ and $\nu3^4$ in $^{45}$Sc. We note here that the configurations fixed for $^{45}$Sc pertain to deformed orbitals; therefore, they
represent much richer correlations than the spherical $f_{7/2}$
configurations usually defined in the context of the shell model. 

Calculations were performed for eight zero-range Skyrme-type
functionals, UNEDF0~\protect\cite{(Kor10c)},
UNEDF1~\protect\cite{(Kor12b)}, SkXc~\protect\cite{(Bro98)},
SIII~\protect\cite{(Bei75b)}, SkM*~\protect\cite{(Bar82c)},
SLy4~\protect\cite{(Cha98a)}, SAMi~\protect\cite{(Roc12b)}, and SkO$^\prime$~\protect\cite{(Rei99)}, and for two finite-range
functionals, D1S~\protect\cite{(Ber91c)} and N$^3$LO
REG6d.190617~\protect\cite{(Ben20)}. The goal of trying several
different variants of functionals was to estimate the order of magnitude
and spread of the results. For all functionals, the experimental value of the electric quadrupole moment of $Q=-0.216(9)$\,b was reached near $Q_{20}=-1$\,b. 

The calculated values of $\mu$ and $\Omega$ strongly depend on several input ingredients of the calculation. First, even at $Q_{20}=0$ these values lie far from the Schmidt~\cite{(Sch37)} and Schwartz~\cite{Schwartz1955} single-particle estimates. This can be attributed to a strong quadrupole coupling to the occupied neutron $f_{7/2}$ orbitals, which decreases both $\mu$ and $\Omega$. Second, the spin polarization, which acts for the Landau spin-spin terms included, also significantly decreases $\mu$ and $\Omega$. Following Ref.~\cite{(Ben02d)}, we
parametrized the spin-spin terms by the standard isoscalar and isovector
Landau parameters $g_0=0.4$ and $g'_0=1.2$, respectively. The value of $g'_0$ was confirmed in global adjustments performed in Ref.~\cite{(Sas21)}, which gave $g'_0=1.0(4)$, 1.3(4), and 1.7(4) for functionals
SkO$^\prime$, SLy4, and UNEDF1, respectively. Third, with increasing intrinsic oblate deformation, both $\mu$ and $\Omega$ increase. The latter effect can be removed by pinning down the intrinsic deformation to the experimental value of $Q$, see stars in Fig.~\ref{fig:M1-M3}. The shaded area in Fig.~\ref{fig:M1-M3} covers the range of results given by all starred points, and thus represents a very rough estimate of the averages and rms deviations of the DFT results: $\mu_{\text{DFT}}=+4.74(6)$\,$\mu_N$ and $\Omega_{\text{DFT}}=+0.245(17)$\,$\mu_N$\,b. 
% The experimental value of $\mu$ lies between those obtained with and without spin polarization. So, in principle, it can be reproduced by adjusting Landau parameters or adjusting effective gyromagnetic factors. 

\subsection{Interpretation}

\begin{figure}[tb]
\begin{center}
\includegraphics[width=\columnwidth]{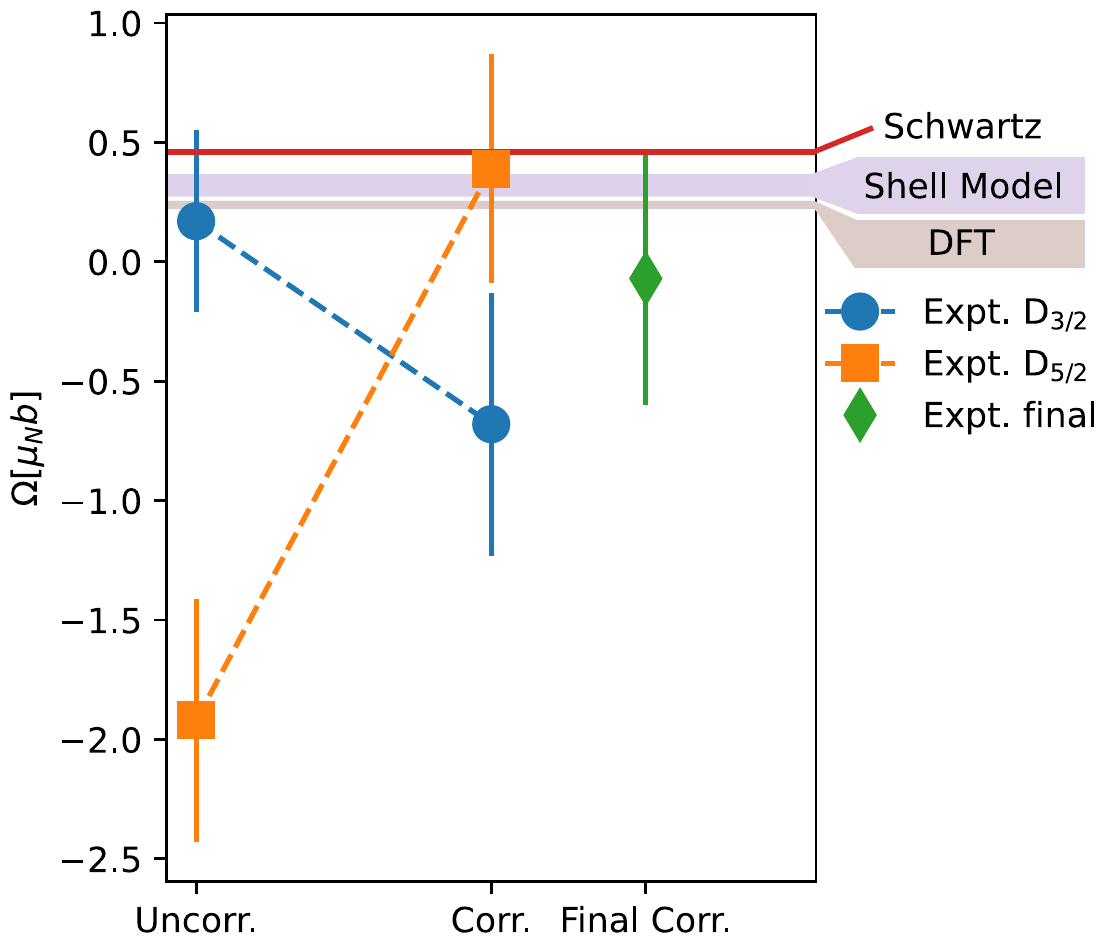}
\caption{Graphical comparison of experimental values of $\Omega$, with and without second-order corrections, and the theoretical predictions. The coloured bands indicate the theoretical uncertainties. 
\label{fig:comparison}
}
\end{center}
\end{figure}

We summarize our experimental and theoretical results in Table~\ref{tab:omega} and graphically in Fig.\ref{fig:comparison}. The inclusion of the second-order shifts brings the extracted value of $\Omega$ obtained for the two different $D_J$ states into reasonable agreement, providing a measure of confidence that these second-order shifts and the values of $C/\Omega$ are calculated accurately. The final value, obtained as the dispersion-corrected weighted mean of the two values, is also shown on the figure, alongside the theoretical values, which are shown as shaded bands. 

The final experimental value agrees well with all theory values. It is interesting to note however that the large-scale shell model and DFT calculations yield smaller values of $\Omega$ than the single-particle Schwartz estimate, bringing these more refined models into closer agreement with experiment. A reduction of the experimental error bar by at least one order of magnitude would be required to provide a more stringent test of the different theoretical approaches.

\section{Conclusion}

We have measured the magnetic octupole moment $\Omega$ in $^{45}$Sc, using a high-precision experimental technique and state-of-the-art atomic calculations. Our shell-model and DFT calculations (with no parameter adjustments) reproduce the values of $\Omega$, of $Q$ up to about 10\%, and of $\mu$ up to 3\%. Further work is required to improve the experimental precision further in order to stringently test nuclear theory. An increase in precision of about a factor of 10 would likely be required to do so, which is out of reach of our current experimental apparatus. A longer rf-interaction region and finer control of the external magnetic field strength would be required. Future experimental work on extending the measurements to other elements, and also to radioactive isotopes, would be very beneficial. This experimental effort should be matched by accurate atomic structure and nuclear structure calculations. As illustrated in this work, atomic and nuclear theory are capable of producing results with sufficient accuracy for such future programs. As a next experimental step, we are currently designing and constructing a collinear RIS laser-RF apparatus which we will use to perform measurements on radioactive isotopes. Candidates for future studies on radioactive isotopes include In and Bi, both having a single proton (hole) outside (inside) of a closed shell, which furthermore feature comparatively larger values of the hyperfine $C$-constant~\cite{Kusch1957,Landman1970}. 

\section{Acknowledgements}

RPDG received funding from the European Union’s Horizon 2020 research and innovation programme under the Marie Sk{\l}odowska-Curie grant agreement No 844829. BKS acknowledges use of Vikram-100 HPC cluster of Physical Research Laboratory, Ahmedabad for atomic calculations. CY acknowledges support of National Natural Science Foundation of China (11775316). This work was supported in part by STFC Grant numbers ST/M006433/1 and ST/P003885/1, and by the Polish National Science Centre under Contract No.\ 2018/31/B/ST2/02220. We acknowledge the CSC-IT Center for Science Ltd., Finland, for the allocation of computational resources. Fruitful discussions with W. Gins and \'A. Koszor\'us are gratefully acknowledged.


\begin{thebibliography}{73}
\expandafter\ifx\csname natexlab\endcsname\relax\def\natexlab#1{#1}\fi
\providecommand{\bibinfo}[2]{#2}
\ifx\xfnm\relax \def\xfnm[#1]{\unskip,\space#1}\fi
%Type = Article
\bibitem[{Neyens et~al.(2005)Neyens, Kowalska, Yordanov, Blaum, Himpe, Lievens,
  Mallion, Neugart, Vermeulen, Utsuno et~al.}]{neyens2005}
\bibinfo{author}{G.~Neyens}, \bibinfo{author}{M.~Kowalska},
  \bibinfo{author}{D.~Yordanov}, \bibinfo{author}{K.~Blaum},
  \bibinfo{author}{P.~Himpe}, \bibinfo{author}{P.~Lievens},
  \bibinfo{author}{S.~Mallion}, \bibinfo{author}{R.~Neugart},
  \bibinfo{author}{N.~Vermeulen}, \bibinfo{author}{Y.~Utsuno}, et~al.,
\newblock \bibinfo{title}{Measurement of the spin and magnetic moment of
  {$^{31}$Mg}: Evidence for a strongly deformed intruder ground state},
\newblock \bibinfo{journal}{Phys. Rev. Lett.} \bibinfo{volume}{94}
  (\bibinfo{year}{2005}) \bibinfo{pages}{022501}.
%Type = Article
\bibitem[{Flanagan et~al.(2009)Flanagan, Vingerhoets, Avgoulea, Billowes,
  Bissell, Blaum, Cheal, De~Rydt, Fedosseev, Forest et~al.}]{flanagan2009}
\bibinfo{author}{K.~Flanagan}, \bibinfo{author}{P.~Vingerhoets},
  \bibinfo{author}{M.~Avgoulea}, \bibinfo{author}{J.~Billowes},
  \bibinfo{author}{M.~Bissell}, \bibinfo{author}{K.~Blaum},
  \bibinfo{author}{B.~Cheal}, \bibinfo{author}{M.~De~Rydt},
  \bibinfo{author}{V.~Fedosseev}, \bibinfo{author}{D.~Forest}, et~al.,
\newblock \bibinfo{title}{Nuclear spins and magnetic moments of
  {$^{71,73,75}$Cu}: Inversion of $\pi$ 2 $p_{3/2}$ and $\pi$ 1 $f_{5/2}$
  levels in {$^{75}$Cu}},
\newblock \bibinfo{journal}{Phys. Rev. Lett.} \bibinfo{volume}{103}
  (\bibinfo{year}{2009}) \bibinfo{pages}{142501}.
%Type = Article
\bibitem[{Ruiz et~al.(2016)Ruiz, Bissell, Blaum, Ekstr{\"o}m, Fr{\"o}mmgen,
  Hagen, Hammen, Hebeler, Holt, Jansen et~al.}]{ruiz2016}
\bibinfo{author}{R.~G. Ruiz}, \bibinfo{author}{M.~Bissell},
  \bibinfo{author}{K.~Blaum}, \bibinfo{author}{A.~Ekstr{\"o}m},
  \bibinfo{author}{N.~Fr{\"o}mmgen}, \bibinfo{author}{G.~Hagen},
  \bibinfo{author}{M.~Hammen}, \bibinfo{author}{K.~Hebeler},
  \bibinfo{author}{J.~Holt}, \bibinfo{author}{G.~Jansen}, et~al.,
\newblock \bibinfo{title}{Unexpectedly large charge radii of neutron-rich
  calcium isotopes},
\newblock \bibinfo{journal}{Nat. Phys.} \bibinfo{volume}{12}
  (\bibinfo{year}{2016}) \bibinfo{pages}{594--598}.
%Type = Article
\bibitem[{Marsh et~al.(2018)Marsh, Goodacre, Sels, Tsunoda, Andel, Andreyev,
  Althubiti, Atanasov, Barzakh, Billowes et~al.}]{marsh2018}
\bibinfo{author}{B.~Marsh}, \bibinfo{author}{T.~D. Goodacre},
  \bibinfo{author}{S.~Sels}, \bibinfo{author}{Y.~Tsunoda},
  \bibinfo{author}{B.~Andel}, \bibinfo{author}{A.~Andreyev},
  \bibinfo{author}{N.~Althubiti}, \bibinfo{author}{D.~Atanasov},
  \bibinfo{author}{A.~Barzakh}, \bibinfo{author}{J.~Billowes}, et~al.,
\newblock \bibinfo{title}{Characterization of the shape-staggering effect in
  mercury nuclei},
\newblock \bibinfo{journal}{Nat. Phys.} \bibinfo{volume}{14}
  (\bibinfo{year}{2018}) \bibinfo{pages}{1163--1167}.
%Type = Article
\bibitem[{Ichikawa(2018)}]{ichikawa2018}
\bibinfo{author}{Y.~Ichikawa},
\newblock \bibinfo{title}{Magnetic moment of isomeric state of $^{75}${Cu}},
\newblock \bibinfo{journal}{Bulletin of the American Physical Society}
  \bibinfo{volume}{63} (\bibinfo{year}{2018}).
%Type = Article
\bibitem[{Miller et~al.(2019)Miller, Minamisono, Klose, Garand, Kujawa, Lantis,
  Liu, Maa{\ss}, Mantica, Nazarewicz et~al.}]{miller2019}
\bibinfo{author}{A.~J. Miller}, \bibinfo{author}{K.~Minamisono},
  \bibinfo{author}{A.~Klose}, \bibinfo{author}{D.~Garand},
  \bibinfo{author}{C.~Kujawa}, \bibinfo{author}{J.~Lantis},
  \bibinfo{author}{Y.~Liu}, \bibinfo{author}{B.~Maa{\ss}},
  \bibinfo{author}{P.~Mantica}, \bibinfo{author}{W.~Nazarewicz}, et~al.,
\newblock \bibinfo{title}{Proton superfluidity and charge radii in proton-rich
  calcium isotopes},
\newblock \bibinfo{journal}{Nat. Phys.} \bibinfo{volume}{15}
  (\bibinfo{year}{2019}) \bibinfo{pages}{432--436}.
%Type = Article
\bibitem[{Campbell et~al.(2016)Campbell, Moore, and Pearson}]{Campbell2016}
\bibinfo{author}{P.~Campbell}, \bibinfo{author}{I.~Moore},
  \bibinfo{author}{M.~Pearson},
\newblock \bibinfo{title}{Laser spectroscopy for nuclear structure physics},
\newblock \bibinfo{journal}{Prog. Part. Nucl. Phys.} \bibinfo{volume}{86}
  (\bibinfo{year}{2016}) \bibinfo{pages}{127 -- 180}.
%Type = Article
\bibitem[{Papoulia et~al.(2016)Papoulia, Carlsson, and Ekman}]{Papoulia2016}
\bibinfo{author}{A.~Papoulia}, \bibinfo{author}{B.~G. Carlsson},
  \bibinfo{author}{J.~Ekman},
\newblock \bibinfo{title}{Effect of realistic nuclear charge distributions on
  isotope shifts and progress towards the extraction of higher-order nuclear
  radial moments},
\newblock \bibinfo{journal}{Phys. Rev. A} \bibinfo{volume}{94}
  (\bibinfo{year}{2016}) \bibinfo{pages}{042502}.
%Type = Article
\bibitem[{Reinhard et~al.(2020)Reinhard, Nazarewicz, and
  Garcia~Ruiz}]{reinhard2020}
\bibinfo{author}{P.-G. Reinhard}, \bibinfo{author}{W.~Nazarewicz},
  \bibinfo{author}{R.~Garcia~Ruiz},
\newblock \bibinfo{title}{Beyond the charge radius: The information content of
  the fourth radial moment},
\newblock \bibinfo{journal}{Phys. Rev. C} \bibinfo{volume}{101}
  (\bibinfo{year}{2020}) \bibinfo{pages}{021301}.
%Type = Article
\bibitem[{Takamine et~al.(2014)Takamine, Wada, Okada, Sonoda, Schury, Nakamura,
  Kanai, Kubo, Katayama, Ohtani, Wollnik, and Schuessler}]{Takamine2014}
\bibinfo{author}{A.~Takamine}, \bibinfo{author}{M.~Wada},
  \bibinfo{author}{K.~Okada}, \bibinfo{author}{T.~Sonoda},
  \bibinfo{author}{P.~Schury}, \bibinfo{author}{T.~Nakamura},
  \bibinfo{author}{Y.~Kanai}, \bibinfo{author}{T.~Kubo},
  \bibinfo{author}{I.~Katayama}, \bibinfo{author}{S.~Ohtani},
  \bibinfo{author}{H.~Wollnik}, \bibinfo{author}{H.~A. Schuessler},
\newblock \bibinfo{title}{Hyperfine structure constant of the neutron halo
  nucleus $^{11}${Be}$^{+}$},
\newblock \bibinfo{journal}{Phys. Rev. Lett.} \bibinfo{volume}{112}
  (\bibinfo{year}{2014}) \bibinfo{pages}{162502}.
%Type = Article
\bibitem[{Papuga et~al.(2014)Papuga, Bissell, Kreim, Barbieri, Blaum, De~Rydt,
  Duguet, Garcia~Ruiz, Heylen, Kowalska, Neugart, Neyens, N\"ortersh\"auser,
  Rajabali, S\'anchez, Smirnova, Som\`a, and Yordanov}]{papuga2014}
\bibinfo{author}{J.~Papuga}, \bibinfo{author}{M.~L. Bissell},
  \bibinfo{author}{K.~Kreim}, \bibinfo{author}{C.~Barbieri},
  \bibinfo{author}{K.~Blaum}, \bibinfo{author}{M.~De~Rydt},
  \bibinfo{author}{T.~Duguet}, \bibinfo{author}{R.~F. Garcia~Ruiz},
  \bibinfo{author}{H.~Heylen}, \bibinfo{author}{M.~Kowalska},
  \bibinfo{author}{R.~Neugart}, \bibinfo{author}{G.~Neyens},
  \bibinfo{author}{W.~N\"ortersh\"auser}, \bibinfo{author}{M.~M. Rajabali},
  \bibinfo{author}{R.~S\'anchez}, \bibinfo{author}{N.~Smirnova},
  \bibinfo{author}{V.~Som\`a}, \bibinfo{author}{D.~T. Yordanov},
\newblock \bibinfo{title}{Shell structure of potassium isotopes deduced from
  their magnetic moments},
\newblock \bibinfo{journal}{Phys. Rev. C} \bibinfo{volume}{90}
  (\bibinfo{year}{2014}) \bibinfo{pages}{034321}.
%Type = Article
\bibitem[{Zhang et~al.(2015)Zhang, Tandecki, Collister, Aubin, Behr, Gomez,
  Gwinner, Orozco, Pearson, Sprouse et~al.}]{zhang2015}
\bibinfo{author}{J.~Zhang}, \bibinfo{author}{M.~Tandecki},
  \bibinfo{author}{R.~Collister}, \bibinfo{author}{S.~Aubin},
  \bibinfo{author}{J.~Behr}, \bibinfo{author}{E.~Gomez},
  \bibinfo{author}{G.~Gwinner}, \bibinfo{author}{L.~Orozco},
  \bibinfo{author}{M.~Pearson}, \bibinfo{author}{G.~Sprouse}, et~al.,
\newblock \bibinfo{title}{Hyperfine anomalies in {Fr}: boundaries of the
  spherical single particle model},
\newblock \bibinfo{journal}{Phys. Rev. Lett.} \bibinfo{volume}{115}
  (\bibinfo{year}{2015}) \bibinfo{pages}{042501}.
%Type = Article
\bibitem[{Schmidt et~al.(2018)Schmidt, Billowes, Bissell, Blaum, Ruiz, Heylen,
  Malbrunot-Ettenauer, Neyens, {N\"ortersh\"auser}, Plunien, Sailer, Shabaev,
  Skripnikov, Tupitsyn, Volotka, and Yang}]{schmidt2018}
\bibinfo{author}{S.~Schmidt}, \bibinfo{author}{J.~Billowes},
  \bibinfo{author}{M.~Bissell}, \bibinfo{author}{K.~Blaum},
  \bibinfo{author}{R.~G. Ruiz}, \bibinfo{author}{H.~Heylen},
  \bibinfo{author}{S.~Malbrunot-Ettenauer}, \bibinfo{author}{G.~Neyens},
  \bibinfo{author}{W.~{N\"ortersh\"auser}}, \bibinfo{author}{G.~Plunien},
  \bibinfo{author}{S.~Sailer}, \bibinfo{author}{V.~Shabaev},
  \bibinfo{author}{L.~Skripnikov}, \bibinfo{author}{I.~Tupitsyn},
  \bibinfo{author}{A.~Volotka}, \bibinfo{author}{X.~Yang},
\newblock \bibinfo{title}{The nuclear magnetic moment of $^{208}${Bi} and its
  relevance for a test of bound-state strong-field {QED}},
\newblock \bibinfo{journal}{Phys. Lett. B} \bibinfo{volume}{779}
  (\bibinfo{year}{2018}) \bibinfo{pages}{324 -- 330}.
%Type = Article
\bibitem[{Persson(2020)}]{Persson2020}
\bibinfo{author}{J.~R. Persson},
\newblock \bibinfo{title}{Hyperfine anomaly in {Eu} isotopes and the
  universiability of the {Moskowitz–Lombardi} formula},
\newblock \bibinfo{journal}{Atoms} \bibinfo{volume}{8} (\bibinfo{year}{2020}).
%Type = Article
\bibitem[{Stroke et~al.(2000)Stroke, Duong, and Pinard}]{stroke2000}
\bibinfo{author}{H.~H. Stroke}, \bibinfo{author}{H.~Duong},
  \bibinfo{author}{J.~Pinard},
\newblock \bibinfo{title}{{Bohr}--{Weisskopf} effect: influence of the
  distributed nuclear magnetization on hfs},
\newblock \bibinfo{journal}{Hyperfine Interact.} \bibinfo{volume}{129}
  (\bibinfo{year}{2000}) \bibinfo{pages}{319--335}.
%Type = Article
\bibitem[{Karpeshin and Trzhaskovskaya(2015)}]{karpeshim2015}
\bibinfo{author}{F.~Karpeshin}, \bibinfo{author}{M.~Trzhaskovskaya},
\newblock \bibinfo{title}{The theory of the bohr–weisskopf effect in the
  hyperfine structure},
\newblock \bibinfo{journal}{Nucl. Phys. A} \bibinfo{volume}{941}
  (\bibinfo{year}{2015}) \bibinfo{pages}{66 -- 77}.
%Type = Article
\bibitem[{Childs(1971)}]{Childs1971}
\bibinfo{author}{W.~J. Childs},
\newblock \bibinfo{title}{Off-diagonal hyperfine structure in $^{45}${Sc}},
\newblock \bibinfo{journal}{Phys. Rev. A} \bibinfo{volume}{4}
  (\bibinfo{year}{1971}) \bibinfo{pages}{1767--1774}.
%Type = Article
\bibitem[{Daly and Holloway(1954)}]{Daly1954}
\bibinfo{author}{R.~T. Daly}, \bibinfo{author}{J.~H. Holloway},
\newblock \bibinfo{title}{Nuclear magnetic octupole moments of the stable
  gallium isotopes},
\newblock \bibinfo{journal}{Phys. Rev.} \bibinfo{volume}{96}
  (\bibinfo{year}{1954}) \bibinfo{pages}{539--540}.
%Type = Article
\bibitem[{Brown and King(1966)}]{Brown1966}
\bibinfo{author}{H.~H. Brown}, \bibinfo{author}{J.~G. King},
\newblock \bibinfo{title}{Hyperfine structure and octopole interaction in
  stable bromine isotopes},
\newblock \bibinfo{journal}{Phys. Rev.} \bibinfo{volume}{142}
  (\bibinfo{year}{1966}) \bibinfo{pages}{53--59}.
%Type = Article
\bibitem[{Faust and Chow~Chiu(1963)}]{Faust1963}
\bibinfo{author}{W.~L. Faust}, \bibinfo{author}{L.~Y. Chow~Chiu},
\newblock \bibinfo{title}{Hyperfine structure of the metastable
  ${(4p)}^{5}(5s)^{3}{P}_{2}$ state of $_{36}\mathrm{Kr}^{83}$},
\newblock \bibinfo{journal}{Phys. Rev.} \bibinfo{volume}{129}
  (\bibinfo{year}{1963}) \bibinfo{pages}{1214--1220}.
%Type = Article
\bibitem[{Eck and Kusch(1957)}]{Kusch1957}
\bibinfo{author}{T.~G. Eck}, \bibinfo{author}{P.~Kusch},
\newblock \bibinfo{title}{Hfs of the $5^{2}p_{\frac{3}{2}}$ state of
  {$^{115}$In} and {$^{113}$In}: Octupole interactions in the stable isotopes
  of indium},
\newblock \bibinfo{journal}{Phys. Rev.} \bibinfo{volume}{106}
  (\bibinfo{year}{1957}) \bibinfo{pages}{958--964}.
%Type = Article
\bibitem[{Jaccarino et~al.(1954)Jaccarino, King, Satten, and
  Stroke}]{Jaccarino1954}
\bibinfo{author}{V.~Jaccarino}, \bibinfo{author}{J.~G. King},
  \bibinfo{author}{R.~A. Satten}, \bibinfo{author}{H.~H. Stroke},
\newblock \bibinfo{title}{Hyperfine structure of {I}$^{127}$. nuclear magnetic
  octupole moment},
\newblock \bibinfo{journal}{Phys. Rev.} \bibinfo{volume}{94}
  (\bibinfo{year}{1954}) \bibinfo{pages}{1798--1799}.
%Type = Article
\bibitem[{Faust and McDermott(1961)}]{Faust1961}
\bibinfo{author}{W.~L. Faust}, \bibinfo{author}{M.~N. McDermott},
\newblock \bibinfo{title}{Hyperfine structure of the ${(5p)}^{5}(6s)^{3}p_{2}$
  state of $_{54}\mathrm{Xe}^{129}$ and $_{54}\mathrm{Xe}^{131}$},
\newblock \bibinfo{journal}{Phys. Rev.} \bibinfo{volume}{123}
  (\bibinfo{year}{1961}) \bibinfo{pages}{198--204}.
%Type = Article
\bibitem[{Gerginov et~al.(2003)Gerginov, Derevianko, and Tanner}]{Gerginov2003}
\bibinfo{author}{V.~Gerginov}, \bibinfo{author}{A.~Derevianko},
  \bibinfo{author}{C.~E. Tanner},
\newblock \bibinfo{title}{Observation of the nuclear magnetic octupole moment
  of $^{133}\mathrm{C}\mathrm{s}$},
\newblock \bibinfo{journal}{Phys. Rev. Lett.} \bibinfo{volume}{91}
  (\bibinfo{year}{2003}) \bibinfo{pages}{072501}.
%Type = Article
\bibitem[{Lewty et~al.(2012)Lewty, Chuah, Cazan, Sahoo, and
  Barrett}]{Lewty2012}
\bibinfo{author}{N.~C. Lewty}, \bibinfo{author}{B.~L. Chuah},
  \bibinfo{author}{R.~Cazan}, \bibinfo{author}{B.~K. Sahoo},
  \bibinfo{author}{M.~D. Barrett},
\newblock \bibinfo{title}{Spectroscopy on a single trapped $^{137}$ba$+$ ion
  for nuclear magnetic octupole moment determination},
\newblock \bibinfo{journal}{Opt. Express} \bibinfo{volume}{20}
  (\bibinfo{year}{2012}) \bibinfo{pages}{21379--21384}.
%Type = Article
\bibitem[{Unsworth(1969)}]{Unsworth1969}
\bibinfo{author}{P.~J. Unsworth},
\newblock \bibinfo{title}{Nuclear dipole, quadrupole and octupole moments of
  $^{155}${Gd} by atomic beam magnetic resonance},
\newblock \bibinfo{journal}{J. Phys. B} \bibinfo{volume}{2}
  (\bibinfo{year}{1969}) \bibinfo{pages}{122--133}.
%Type = Article
\bibitem[{Singh et~al.(2013)Singh, Angom, and Natarajan}]{Singh2013}
\bibinfo{author}{A.~K. Singh}, \bibinfo{author}{D.~Angom},
  \bibinfo{author}{V.~Natarajan},
\newblock \bibinfo{title}{Observation of the nuclear magnetic octupole moment
  of ${}^{173}${Yb} from precise measurements of the hyperfine structure in the
  ${{}^{3}P}_{2}$ state},
\newblock \bibinfo{journal}{Phys. Rev. A} \bibinfo{volume}{87}
  (\bibinfo{year}{2013}) \bibinfo{pages}{012512}.
%Type = Article
\bibitem[{McDermott and Lichten(1960)}]{McDermott1960}
\bibinfo{author}{M.~N. McDermott}, \bibinfo{author}{W.~L. Lichten},
\newblock \bibinfo{title}{Hyperfine structure of the $6^{3}p_{2}$ state of
  $_{80}\mathrm{Hg}^{199}$ and $_{80}\mathrm{Hg}^{201}$. properties of
  metastable states of mercury},
\newblock \bibinfo{journal}{Phys. Rev.} \bibinfo{volume}{119}
  (\bibinfo{year}{1960}) \bibinfo{pages}{134--143}.
%Type = Article
\bibitem[{Landman and Lurio(1970)}]{Landman1970}
\bibinfo{author}{D.~A. Landman}, \bibinfo{author}{A.~Lurio},
\newblock \bibinfo{title}{Hyperfine structure of the ${(6p)}^{3}$ configuration
  of {Bi}$^{209}$},
\newblock \bibinfo{journal}{Phys. Rev. A} \bibinfo{volume}{1}
  (\bibinfo{year}{1970}) \bibinfo{pages}{1330--1338}.
%Type = Article
\bibitem[{Fuller(1976)}]{fuller1976}
\bibinfo{author}{G.~H. Fuller},
\newblock \bibinfo{title}{Nuclear spins and moments},
\newblock \bibinfo{journal}{J. Phys. Chem. Ref. Data} \bibinfo{volume}{5}
  (\bibinfo{year}{1976}) \bibinfo{pages}{835--1092}.
%Type = Article
\bibitem[{Schwartz(1955)}]{Schwartz1955}
\bibinfo{author}{C.~Schwartz},
\newblock \bibinfo{title}{Theory of hyperfine structure},
\newblock \bibinfo{journal}{Phys. Rev.} \bibinfo{volume}{97}
  (\bibinfo{year}{1955}) \bibinfo{pages}{380--395}.
%Type = Article
\bibitem[{de~Groote et~al.(2021)de~Groote, Kujanp{\"a}{\"a}, Koszor{\'u}s, Li,
  and Moore}]{degroote2021}
\bibinfo{author}{R.~de~Groote}, \bibinfo{author}{S.~Kujanp{\"a}{\"a}},
  \bibinfo{author}{{\'A}.~Koszor{\'u}s}, \bibinfo{author}{J.~Li},
  \bibinfo{author}{I.~Moore},
\newblock \bibinfo{title}{Magnetic octupole moment of $^{173}${Yb} using
  collinear laser spectroscopy},
\newblock \bibinfo{journal}{Physical Review A} \bibinfo{volume}{103}
  (\bibinfo{year}{2021}) \bibinfo{pages}{032826}.
%Type = Article
\bibitem[{de~Groote et~al.(2019)de~Groote, Billowes, Binnersley, Bissell,
  Cocolios, Goodacre, Farooq-Smith, Fedorov, Flanagan, Franchoo
  et~al.}]{degroote2019}
\bibinfo{author}{R.~de~Groote}, \bibinfo{author}{J.~Billowes},
  \bibinfo{author}{C.~Binnersley}, \bibinfo{author}{M.~Bissell},
  \bibinfo{author}{T.~Cocolios}, \bibinfo{author}{T.~D. Goodacre},
  \bibinfo{author}{G.~Farooq-Smith}, \bibinfo{author}{D.~Fedorov},
  \bibinfo{author}{K.~Flanagan}, \bibinfo{author}{S.~Franchoo}, et~al.,
\newblock \bibinfo{title}{Precise measurement and microscopic description of
  charge radii of exotic copper isotopes: global trends and odd-even
  variations},
\newblock \bibinfo{journal}{arXiv preprint arXiv:1911.08765}
  (\bibinfo{year}{2019}).
%Type = Article
\bibitem[{Reponen et~al.(2021)Reponen, de~Groote, Al~Ayoubi, Beliuskina,
  Bissell, Campbell, Ca{\~n}ete, Cheal, Chrysalidis, Delafosse
  et~al.}]{reponen2021}
\bibinfo{author}{M.~Reponen}, \bibinfo{author}{R.~de~Groote},
  \bibinfo{author}{L.~Al~Ayoubi}, \bibinfo{author}{O.~Beliuskina},
  \bibinfo{author}{M.~Bissell}, \bibinfo{author}{P.~Campbell},
  \bibinfo{author}{L.~Ca{\~n}ete}, \bibinfo{author}{B.~Cheal},
  \bibinfo{author}{K.~Chrysalidis}, \bibinfo{author}{C.~Delafosse}, et~al.,
\newblock \bibinfo{title}{Evidence of a sudden increase in the nuclear size of
  proton-rich silver-96},
\newblock \bibinfo{journal}{Nature Communications} \bibinfo{volume}{12}
  (\bibinfo{year}{2021}) \bibinfo{pages}{1--8}.
%Type = Article
\bibitem[{Childs(1992)}]{Childs1992}
\bibinfo{author}{W.~Childs},
\newblock \bibinfo{title}{Overview of laser-radiofrequency double-resonance
  studies of atomic, molecular, and ionic beams},
\newblock \bibinfo{journal}{Physics reports} \bibinfo{volume}{211}
  (\bibinfo{year}{1992}) \bibinfo{pages}{113--165}.
%Type = Article
\bibitem[{Vernon et~al.(2019)Vernon, Billowes, Binnersley, Bissell, Cocolios,
  Farooq-Smith, Flanagan, Ruiz, Gins, de~Groote, Koszor{\'u}s, Lynch, Neyens,
  Ricketts, Wendt, Wilkins, and Yang}]{Vernon2019}
\bibinfo{author}{A.~Vernon}, \bibinfo{author}{J.~Billowes},
  \bibinfo{author}{C.~Binnersley}, \bibinfo{author}{M.~Bissell},
  \bibinfo{author}{T.~Cocolios}, \bibinfo{author}{G.~Farooq-Smith},
  \bibinfo{author}{K.~Flanagan}, \bibinfo{author}{R.~G. Ruiz},
  \bibinfo{author}{W.~Gins}, \bibinfo{author}{R.~de~Groote},
  \bibinfo{author}{{\'A}.~Koszor{\'u}s}, \bibinfo{author}{K.~Lynch},
  \bibinfo{author}{G.~Neyens}, \bibinfo{author}{C.~Ricketts},
  \bibinfo{author}{K.~Wendt}, \bibinfo{author}{S.~Wilkins},
  \bibinfo{author}{X.~Yang},
\newblock \bibinfo{title}{Simulation of the relative atomic populations of
  elements $1<z<89$ following charge exchange tested with collinear resonance
  ionization spectroscopy of indium},
\newblock \bibinfo{journal}{Spectrochimica Acta Part B: Atomic Spectroscopy}
  \bibinfo{volume}{153} (\bibinfo{year}{2019}) \bibinfo{pages}{61 -- 83}.
%Type = Article
\bibitem[{Nielsen et~al.(1983)Nielsen, Poulsen, Thorsen, and
  Crosswhite}]{Nielsen1983}
\bibinfo{author}{U.~Nielsen}, \bibinfo{author}{O.~Poulsen},
  \bibinfo{author}{P.~Thorsen}, \bibinfo{author}{H.~Crosswhite},
\newblock \bibinfo{title}{Collinear laser-rf double-resonance spectroscopy:
  {U}-235 {II} hyperfine structure},
\newblock \bibinfo{journal}{Phys. Rev. Lett.} \bibinfo{volume}{51}
  (\bibinfo{year}{1983}) \bibinfo{pages}{1749}.
%Type = Article
\bibitem[{Steppenbeck et~al.(2013)Steppenbeck, Takeuchi, Aoi, Doornenbal,
  Matsushita, Wang, Baba, Fukuda, Go, Honma et~al.}]{steppenbeck2013}
\bibinfo{author}{D.~Steppenbeck}, \bibinfo{author}{S.~Takeuchi},
  \bibinfo{author}{N.~Aoi}, \bibinfo{author}{P.~Doornenbal},
  \bibinfo{author}{M.~Matsushita}, \bibinfo{author}{H.~Wang},
  \bibinfo{author}{H.~Baba}, \bibinfo{author}{N.~Fukuda},
  \bibinfo{author}{S.~Go}, \bibinfo{author}{M.~Honma}, et~al.,
\newblock \bibinfo{title}{Evidence for a new nuclear 'magic number' from the
  level structure of $^{54}${Ca}},
\newblock \bibinfo{journal}{Nature} \bibinfo{volume}{502}
  (\bibinfo{year}{2013}) \bibinfo{pages}{207--210}.
%Type = Article
\bibitem[{Wienholtz et~al.(2013)Wienholtz, Beck, Blaum, Borgmann, Breitenfeldt,
  Cakirli, George, Herfurth, Holt, Kowalska et~al.}]{wienholtz2013}
\bibinfo{author}{F.~Wienholtz}, \bibinfo{author}{D.~Beck},
  \bibinfo{author}{K.~Blaum}, \bibinfo{author}{C.~Borgmann},
  \bibinfo{author}{M.~Breitenfeldt}, \bibinfo{author}{R.~B. Cakirli},
  \bibinfo{author}{S.~George}, \bibinfo{author}{F.~Herfurth},
  \bibinfo{author}{J.~Holt}, \bibinfo{author}{M.~Kowalska}, et~al.,
\newblock \bibinfo{title}{Masses of exotic calcium isotopes pin down nuclear
  forces},
\newblock \bibinfo{journal}{Nature} \bibinfo{volume}{498}
  (\bibinfo{year}{2013}) \bibinfo{pages}{346--349}.
%Type = Article
\bibitem[{Tanaka et~al.(2020)Tanaka, Takechi, Homma, Fukuda, Nishimura, Suzuki,
  Tanaka, Moriguchi, Ahn, Aimaganbetov, Amano, Arakawa, Bagchi, Behr,
  Burtebayev, Chikaato, Du, Ebata, Fujii, Fukuda, Geissel, Hori, Horiuchi,
  Hoshino, Igosawa, Ikeda, Inabe, Inomata, Itahashi, Izumikawa, Kamioka, Kanda,
  Kato, Kenzhina, Korkulu, Kuk, Kusaka, Matsuta, Mihara, Miyata, Nagae,
  Nakamura, Nassurlla, Nishimuro, Nishizuka, Ohnishi, Ohtake, Ohtsubo, Omika,
  Ong, Ozawa, Prochazka, Sakurai, Scheidenberger, Shimizu, Sugihara, Sumikama,
  Suzuki, Suzuki, Takeda, Tanaka, Tanihata, Wada, Wakayama, Yagi, Yamaguchi,
  Yanagihara, Yanagisawa, Yoshida, and Zholdybayev}]{Tanaka2020}
\bibinfo{author}{M.~Tanaka}, \bibinfo{author}{M.~Takechi},
  \bibinfo{author}{A.~Homma}, \bibinfo{author}{M.~Fukuda},
  \bibinfo{author}{D.~Nishimura}, \bibinfo{author}{T.~Suzuki},
  \bibinfo{author}{Y.~Tanaka}, \bibinfo{author}{T.~Moriguchi},
  \bibinfo{author}{D.~S. Ahn}, \bibinfo{author}{A.~Aimaganbetov},
  \bibinfo{author}{M.~Amano}, \bibinfo{author}{H.~Arakawa},
  \bibinfo{author}{S.~Bagchi}, \bibinfo{author}{K.-H. Behr},
  \bibinfo{author}{N.~Burtebayev}, \bibinfo{author}{K.~Chikaato},
  \bibinfo{author}{H.~Du}, \bibinfo{author}{S.~Ebata},
  \bibinfo{author}{T.~Fujii}, \bibinfo{author}{N.~Fukuda},
  \bibinfo{author}{H.~Geissel}, \bibinfo{author}{T.~Hori},
  \bibinfo{author}{W.~Horiuchi}, \bibinfo{author}{S.~Hoshino},
  \bibinfo{author}{R.~Igosawa}, \bibinfo{author}{A.~Ikeda},
  \bibinfo{author}{N.~Inabe}, \bibinfo{author}{K.~Inomata},
  \bibinfo{author}{K.~Itahashi}, \bibinfo{author}{T.~Izumikawa},
  \bibinfo{author}{D.~Kamioka}, \bibinfo{author}{N.~Kanda},
  \bibinfo{author}{I.~Kato}, \bibinfo{author}{I.~Kenzhina},
  \bibinfo{author}{Z.~Korkulu}, \bibinfo{author}{Y.~Kuk},
  \bibinfo{author}{K.~Kusaka}, \bibinfo{author}{K.~Matsuta},
  \bibinfo{author}{M.~Mihara}, \bibinfo{author}{E.~Miyata},
  \bibinfo{author}{D.~Nagae}, \bibinfo{author}{S.~Nakamura},
  \bibinfo{author}{M.~Nassurlla}, \bibinfo{author}{K.~Nishimuro},
  \bibinfo{author}{K.~Nishizuka}, \bibinfo{author}{K.~Ohnishi},
  \bibinfo{author}{M.~Ohtake}, \bibinfo{author}{T.~Ohtsubo},
  \bibinfo{author}{S.~Omika}, \bibinfo{author}{H.~J. Ong},
  \bibinfo{author}{A.~Ozawa}, \bibinfo{author}{A.~Prochazka},
  \bibinfo{author}{H.~Sakurai}, \bibinfo{author}{C.~Scheidenberger},
  \bibinfo{author}{Y.~Shimizu}, \bibinfo{author}{T.~Sugihara},
  \bibinfo{author}{T.~Sumikama}, \bibinfo{author}{H.~Suzuki},
  \bibinfo{author}{S.~Suzuki}, \bibinfo{author}{H.~Takeda},
  \bibinfo{author}{Y.~K. Tanaka}, \bibinfo{author}{I.~Tanihata},
  \bibinfo{author}{T.~Wada}, \bibinfo{author}{K.~Wakayama},
  \bibinfo{author}{S.~Yagi}, \bibinfo{author}{T.~Yamaguchi},
  \bibinfo{author}{R.~Yanagihara}, \bibinfo{author}{Y.~Yanagisawa},
  \bibinfo{author}{K.~Yoshida}, \bibinfo{author}{T.~K. Zholdybayev},
\newblock \bibinfo{title}{Swelling of doubly magic $^{48}${Ca} core in {Ca}
  isotopes beyond $n=28$},
\newblock \bibinfo{journal}{Phys. Rev. Lett.} \bibinfo{volume}{124}
  (\bibinfo{year}{2020}) \bibinfo{pages}{102501}.
%Type = Article
\bibitem[{Achakovskiy et~al.(2014)Achakovskiy, Kamerdzhiev, Saperstein, and
  Tolokonnikov}]{(Ach14)}
\bibinfo{author}{O.~I. Achakovskiy}, \bibinfo{author}{S.~P. Kamerdzhiev},
  \bibinfo{author}{E.~E. Saperstein}, \bibinfo{author}{S.~V. Tolokonnikov},
\newblock \bibinfo{title}{Magnetic moments of odd-odd spherical nuclei},
\newblock \bibinfo{journal}{The European Physical Journal A}
  \bibinfo{volume}{50} (\bibinfo{year}{2014}) \bibinfo{pages}{6}.
%Type = Article
\bibitem[{Bonneau et~al.(2015)Bonneau, Minkov, Duc, Quentin, and
  Bartel}]{(Bon15)}
\bibinfo{author}{L.~Bonneau}, \bibinfo{author}{N.~Minkov},
  \bibinfo{author}{D.~D. Duc}, \bibinfo{author}{P.~Quentin},
  \bibinfo{author}{J.~Bartel},
\newblock \bibinfo{title}{Effect of core polarization on magnetic dipole
  moments in deformed odd-mass nuclei},
\newblock \bibinfo{journal}{Phys. Rev. C} \bibinfo{volume}{91}
  (\bibinfo{year}{2015}) \bibinfo{pages}{054307}.
%Type = Article
\bibitem[{Borrajo and Egido(2017)}]{(Bor17)}
\bibinfo{author}{M.~Borrajo}, \bibinfo{author}{J.~L. Egido},
\newblock \bibinfo{title}{Ground-state properties of even and odd {Magnesium}
  isotopes in a symmetry-conserving approach},
\newblock \bibinfo{journal}{Physics Letters B} \bibinfo{volume}{764}
  (\bibinfo{year}{2017}) \bibinfo{pages}{328 -- 334}.
%Type = Article
\bibitem[{Brown et~al.(1980)Brown, Chung, and Wildenthal}]{brown1980}
\bibinfo{author}{B.~A. Brown}, \bibinfo{author}{W.~Chung},
  \bibinfo{author}{B.~Wildenthal},
\newblock \bibinfo{title}{Electromagnetic multipole moments of ground states of
  stable odd-mass nuclei in the sd shell},
\newblock \bibinfo{journal}{Phys. Rev. C} \bibinfo{volume}{22}
  (\bibinfo{year}{1980}) \bibinfo{pages}{774}.
%Type = Article
\bibitem[{Sen'kov and Dmitriev(2002)}]{senkov2002}
\bibinfo{author}{R.~Sen'kov}, \bibinfo{author}{V.~Dmitriev},
\newblock \bibinfo{title}{Nuclear magnetization distribution and hyperfine
  splitting in {Bi}$^{82+}$ ion},
\newblock \bibinfo{journal}{Nucl. Phys. A} \bibinfo{volume}{706}
  (\bibinfo{year}{2002}) \bibinfo{pages}{351 -- 364}.
%Type = Article
\bibitem[{Raeder et~al.(2013)Raeder, Dombsky, Heggen, Lassen, Quenzel,
  Sj{\"o}din, Teigelh{\"o}fer, and Wendt}]{raeder2013source}
\bibinfo{author}{S.~Raeder}, \bibinfo{author}{M.~Dombsky},
  \bibinfo{author}{H.~Heggen}, \bibinfo{author}{J.~Lassen},
  \bibinfo{author}{T.~Quenzel}, \bibinfo{author}{M.~Sj{\"o}din},
  \bibinfo{author}{A.~Teigelh{\"o}fer}, \bibinfo{author}{K.~Wendt},
\newblock \bibinfo{title}{In-source laser spectroscopy developments at
  {TRILIS}—towards spectroscopy on actinium and scandium},
\newblock \bibinfo{journal}{Hyperfine Interact.} \bibinfo{volume}{216}
  (\bibinfo{year}{2013}) \bibinfo{pages}{33--39}.
%Type = Book
\bibitem[{Shavitt and Bartlett(2009)}]{shavitt2009}
\bibinfo{author}{I.~Shavitt}, \bibinfo{author}{R.~J. Bartlett},
  \bibinfo{title}{Many-body methods in chemistry and physics: MBPT and
  coupled-cluster theory}, \bibinfo{publisher}{Cambridge university press},
  \bibinfo{year}{2009}.
%Type = Article
\bibitem[{Sahoo et~al.(2005)Sahoo, Beier, Das, Chaudhuri, and
  Mukherjee}]{sahoo2005}
\bibinfo{author}{B.~K. Sahoo}, \bibinfo{author}{T.~Beier},
  \bibinfo{author}{B.~Das}, \bibinfo{author}{R.~Chaudhuri},
  \bibinfo{author}{D.~Mukherjee},
\newblock \bibinfo{title}{Electron correlation effects in hyperfine
  interactions in $^{45}${Sc} and $^{89}${Y}},
\newblock \bibinfo{journal}{J. Phys. B} \bibinfo{volume}{38}
  (\bibinfo{year}{2005}) \bibinfo{pages}{4379}.
%Type = Techreport
\bibitem[{Stone(2019)}]{stone2019}
\bibinfo{author}{N.~Stone}, \bibinfo{title}{Table of recommended nuclear
  magnetic dipole moments}, \bibinfo{type}{Technical Report}, International
  Atomic Energy Agency, \bibinfo{year}{2019}.
%Type = Article
\bibitem[{Stone(2016)}]{stone2016}
\bibinfo{author}{N.~Stone},
\newblock \bibinfo{title}{Table of nuclear electric quadrupole moments},
\newblock \bibinfo{journal}{Atomic Data and Nuclear Data Tables}
  \bibinfo{volume}{111} (\bibinfo{year}{2016}) \bibinfo{pages}{1--28}.
%Type = Article
\bibitem[{Sahoo(2015)}]{sahoo2015}
\bibinfo{author}{B.~K. Sahoo},
\newblock \bibinfo{title}{Appraising nuclear-octupole-moment contributions to
  the hyperfine structures in $^{211}\mathrm{Fr}$},
\newblock \bibinfo{journal}{Phys. Rev. A} \bibinfo{volume}{92}
  (\bibinfo{year}{2015}) \bibinfo{pages}{052506}.
%Type = Article
\bibitem[{Honma et~al.(2002)Honma, Otsuka, Brown, and
  Mizusaki}]{PhysRevC.65.061301}
\bibinfo{author}{M.~Honma}, \bibinfo{author}{T.~Otsuka}, \bibinfo{author}{B.~A.
  Brown}, \bibinfo{author}{T.~Mizusaki},
\newblock \bibinfo{title}{Effective interaction for pf-shell nuclei},
\newblock \bibinfo{journal}{Phys. Rev. C} \bibinfo{volume}{65}
  (\bibinfo{year}{2002}) \bibinfo{pages}{061301}.
%Type = Article
\bibitem[{Honma et~al.(2005)Honma, Otsuka, Brown, and
  Mizusaki}]{EurPhysJA.25.499}
\bibinfo{author}{M.~Honma}, \bibinfo{author}{T.~Otsuka}, \bibinfo{author}{B.~A.
  Brown}, \bibinfo{author}{T.~Mizusaki},
\newblock \bibinfo{title}{Shell-model description of neutron-rich pf-shell
  nuclei with a new effective interaction {GXPF}1},
\newblock \bibinfo{journal}{Eur. Phys. J. A} \bibinfo{volume}{25}
  (\bibinfo{year}{2005}) \bibinfo{pages}{499}.
%Type = Article
\bibitem[{Abzouzi et~al.(1991)Abzouzi, Caurier, and
  Zuker}]{PhysRevLett.66.1134}
\bibinfo{author}{A.~Abzouzi}, \bibinfo{author}{E.~Caurier},
  \bibinfo{author}{A.~P. Zuker},
\newblock \bibinfo{title}{Influence of saturation properties on shell-model
  calculations},
\newblock \bibinfo{journal}{Phys. Rev. Lett.} \bibinfo{volume}{66}
  (\bibinfo{year}{1991}) \bibinfo{pages}{1134--1137}.
%Type = Article
\bibitem[{Poves et~al.(2001)Poves, Sánchez-Solano, Caurier, and
  Nowacki}]{POVES2001157}
\bibinfo{author}{A.~Poves}, \bibinfo{author}{J.~Sánchez-Solano},
  \bibinfo{author}{E.~Caurier}, \bibinfo{author}{F.~Nowacki},
\newblock \bibinfo{title}{Shell model study of the isobaric chains {A=50, A=51}
  and {A=52}},
\newblock \bibinfo{journal}{Nucl. Phys. A} \bibinfo{volume}{694}
  (\bibinfo{year}{2001}) \bibinfo{pages}{157 -- 198}.
%Type = Article
\bibitem[{Shimizu et~al.(2019)Shimizu, Mizusaki, Utsuno, and
  Tsunoda}]{SHIMIZU2019372}
\bibinfo{author}{N.~Shimizu}, \bibinfo{author}{T.~Mizusaki},
  \bibinfo{author}{Y.~Utsuno}, \bibinfo{author}{Y.~Tsunoda},
\newblock \bibinfo{title}{Thick-restart block lanczos method for large-scale
  shell-model calculations},
\newblock \bibinfo{journal}{Computer Physics Communications}
  \bibinfo{volume}{244} (\bibinfo{year}{2019}) \bibinfo{pages}{372 -- 384}.
%Type = Article
\bibitem[{Utsuno et~al.(1999)Utsuno, Otsuka, Mizusaki, and
  Honma}]{PhysRevC.60.054315}
\bibinfo{author}{Y.~Utsuno}, \bibinfo{author}{T.~Otsuka},
  \bibinfo{author}{T.~Mizusaki}, \bibinfo{author}{M.~Honma},
\newblock \bibinfo{title}{Varying shell gap and deformation in
  {N}$\ensuremath{\sim}20$ unstable nuclei studied by the monte carlo shell
  model},
\newblock \bibinfo{journal}{Phys. Rev. C} \bibinfo{volume}{60}
  (\bibinfo{year}{1999}) \bibinfo{pages}{054315}.
%Type = Article
\bibitem[{Utsuno et~al.(2012)Utsuno, Otsuka, Brown, Honma, Mizusaki, and
  Shimizu}]{PhysRevC.86.051301}
\bibinfo{author}{Y.~Utsuno}, \bibinfo{author}{T.~Otsuka},
  \bibinfo{author}{B.~A. Brown}, \bibinfo{author}{M.~Honma},
  \bibinfo{author}{T.~Mizusaki}, \bibinfo{author}{N.~Shimizu},
\newblock \bibinfo{title}{Shape transitions in exotic si and s isotopes and
  tensor-force-driven {Jahn-Teller} effect},
\newblock \bibinfo{journal}{Phys. Rev. C} \bibinfo{volume}{86}
  (\bibinfo{year}{2012}) \bibinfo{pages}{051301}.
%Type = Article
\bibitem[{Nowacki and Poves(2009)}]{PhysRevC.79.014310}
\bibinfo{author}{F.~Nowacki}, \bibinfo{author}{A.~Poves},
\newblock \bibinfo{title}{New effective interaction for $0\hbar\omega$
  shell-model calculations in the $\mathit{sd}\text{\ensuremath{-}}\mathit{pf}$
  valence space},
\newblock \bibinfo{journal}{Phys. Rev. C} \bibinfo{volume}{79}
  (\bibinfo{year}{2009}) \bibinfo{pages}{014310}.
%Type = Article
\bibitem[{Dobaczewski et~al.(2021)Dobaczewski, B\c{a}czyk, Becker, Bender,
  Bennaceur, Bonnard, Gao, Idini, Konieczka, Kortelainen, Pr{\'o}chniak,
  Romero, Satu{\l}a, Shi, Yu, and Werner}]{(Dob21f)}
\bibinfo{author}{J.~Dobaczewski}, \bibinfo{author}{P.~B\c{a}czyk},
  \bibinfo{author}{P.~Becker}, \bibinfo{author}{M.~Bender},
  \bibinfo{author}{K.~Bennaceur}, \bibinfo{author}{J.~Bonnard},
  \bibinfo{author}{Y.~Gao}, \bibinfo{author}{A.~Idini},
  \bibinfo{author}{M.~Konieczka}, \bibinfo{author}{M.~Kortelainen},
  \bibinfo{author}{L.~Pr{\'o}chniak}, \bibinfo{author}{A.~M. Romero},
  \bibinfo{author}{W.~Satu{\l}a}, \bibinfo{author}{Y.~Shi},
  \bibinfo{author}{L.~F. Yu}, \bibinfo{author}{T.~R. Werner},
\newblock \bibinfo{title}{Solution of universal nonrelativistic nuclear {DFT}
  equations in the {Cartesian} deformed harmonic-oscillator basis. {(IX) {\sc
  hfodd}} (v3.06h): a new version of the program},
\newblock \bibinfo{journal}{J. Phys. G: Nucl. Part. Phys.} \bibinfo{volume}{48}
  (\bibinfo{year}{2021}) \bibinfo{pages}{102001}.
%Type = Article
\bibitem[{Kortelainen et~al.(2010)Kortelainen, Lesinski, Mor\'e, Nazarewicz,
  Sarich, Schunck, Stoitsov, and Wild}]{(Kor10c)}
\bibinfo{author}{M.~Kortelainen}, \bibinfo{author}{T.~Lesinski},
  \bibinfo{author}{J.~Mor\'e}, \bibinfo{author}{W.~Nazarewicz},
  \bibinfo{author}{J.~Sarich}, \bibinfo{author}{N.~Schunck},
  \bibinfo{author}{M.~V. Stoitsov}, \bibinfo{author}{S.~Wild},
\newblock \bibinfo{title}{Nuclear energy density optimization},
\newblock \bibinfo{journal}{Phys. Rev. C} \bibinfo{volume}{82}
  (\bibinfo{year}{2010}) \bibinfo{pages}{024313}.
%Type = Article
\bibitem[{Kortelainen et~al.(2012)Kortelainen, McDonnell, Nazarewicz, Reinhard,
  Sarich, Schunck, Stoitsov, and Wild}]{(Kor12b)}
\bibinfo{author}{M.~Kortelainen}, \bibinfo{author}{J.~McDonnell},
  \bibinfo{author}{W.~Nazarewicz}, \bibinfo{author}{P.-G. Reinhard},
  \bibinfo{author}{J.~Sarich}, \bibinfo{author}{N.~Schunck},
  \bibinfo{author}{M.~V. Stoitsov}, \bibinfo{author}{S.~M. Wild},
\newblock \bibinfo{title}{Nuclear energy density optimization: {Large}
  deformations},
\newblock \bibinfo{journal}{Phys. Rev. C} \bibinfo{volume}{85}
  (\bibinfo{year}{2012}) \bibinfo{pages}{024304}.
%Type = Article
\bibitem[{Brown(1998)}]{(Bro98)}
\bibinfo{author}{B.~A. Brown},
\newblock \bibinfo{title}{New skyrme interaction for normal and exotic nuclei},
\newblock \bibinfo{journal}{Phys. Rev. C} \bibinfo{volume}{58}
  (\bibinfo{year}{1998}) \bibinfo{pages}{220--231}.
%Type = Article
\bibitem[{Beiner et~al.(1975)Beiner, Flocard, Giai, and Quentin}]{(Bei75b)}
\bibinfo{author}{M.~Beiner}, \bibinfo{author}{H.~Flocard},
  \bibinfo{author}{N.~V. Giai}, \bibinfo{author}{P.~Quentin},
\newblock \bibinfo{title}{Nuclear ground-state properties and self-consistent
  calculations with the {Skyrme} interaction: {(I). Spherical} description},
\newblock \bibinfo{journal}{Nuclear Physics A} \bibinfo{volume}{238}
  (\bibinfo{year}{1975}) \bibinfo{pages}{29 -- 69}.
%Type = Article
\bibitem[{Bartel et~al.(1982)Bartel, Quentin, Brack, Guet, and
  H{\aa}kansson}]{(Bar82c)}
\bibinfo{author}{J.~Bartel}, \bibinfo{author}{P.~Quentin},
  \bibinfo{author}{M.~Brack}, \bibinfo{author}{C.~Guet}, \bibinfo{author}{H.-B.
  H{\aa}kansson},
\newblock \bibinfo{title}{Towards a better parametrisation of
  \protect{Skyrme-like} effective forces: A critical study of the \protect{SkM}
  force},
\newblock \bibinfo{journal}{Nuclear Physics A} \bibinfo{volume}{386}
  (\bibinfo{year}{1982}) \bibinfo{pages}{79 -- 100}.
%Type = Article
\bibitem[{Chabanat et~al.(1998)Chabanat, Bonche, Haensel, Meyer, and
  Schaeffer}]{(Cha98a)}
\bibinfo{author}{E.~Chabanat}, \bibinfo{author}{P.~Bonche},
  \bibinfo{author}{P.~Haensel}, \bibinfo{author}{J.~Meyer},
  \bibinfo{author}{R.~Schaeffer},
\newblock \bibinfo{title}{A \protect{Skyrme} parametrization from subnuclear to
  neutron star densities \protect{Part II. Nuclei} far from stabilities},
\newblock \bibinfo{journal}{Nuclear Physics A} \bibinfo{volume}{635}
  (\bibinfo{year}{1998}) \bibinfo{pages}{231 -- 256}.
%Type = Article
\bibitem[{Roca-Maza et~al.(2012)Roca-Maza, Col\`o, and Sagawa}]{(Roc12b)}
\bibinfo{author}{X.~Roca-Maza}, \bibinfo{author}{G.~Col\`o},
  \bibinfo{author}{H.~Sagawa},
\newblock \bibinfo{title}{New skyrme interaction with improved spin-isospin
  properties},
\newblock \bibinfo{journal}{Phys. Rev. C} \bibinfo{volume}{86}
  (\bibinfo{year}{2012}) \bibinfo{pages}{031306}.
%Type = Article
\bibitem[{Reinhard(1999)}]{(Rei99)}
\bibinfo{author}{P.-G. Reinhard},
\newblock \bibinfo{title}{Skyrme forces and giant resonances in exotic nuclei},
\newblock \bibinfo{journal}{Nucl. Phys. A} \bibinfo{volume}{649}
  (\bibinfo{year}{1999}) \bibinfo{pages}{305c}.
%Type = Article
\bibitem[{Berger et~al.(1991)Berger, Girod, and Gogny}]{(Ber91c)}
\bibinfo{author}{J.~Berger}, \bibinfo{author}{M.~Girod},
  \bibinfo{author}{D.~Gogny},
\newblock \bibinfo{title}{Time-dependent quantum collective dynamics applied to
  nuclear fission},
\newblock \bibinfo{journal}{Computer Physics Communications}
  \bibinfo{volume}{63} (\bibinfo{year}{1991}) \bibinfo{pages}{365 -- 374}.
%Type = Article
\bibitem[{Bennaceur et~al.(2020)Bennaceur, Dobaczewski, Haverinen, and
  Kortelainen}]{(Ben20)}
\bibinfo{author}{K.~Bennaceur}, \bibinfo{author}{J.~Dobaczewski},
  \bibinfo{author}{T.~Haverinen}, \bibinfo{author}{M.~Kortelainen},
\newblock \bibinfo{title}{Properties of spherical and deformed nuclei using
  regularized pseudopotentials in nuclear {DFT}},
\newblock \bibinfo{journal}{Journal of Physics G: Nuclear and Particle Physics}
  \bibinfo{volume}{47} (\bibinfo{year}{2020}) \bibinfo{pages}{105101}.
%Type = Article
\bibitem[{Schmidt(1937)}]{(Sch37)}
\bibinfo{author}{T.~Schmidt},
\newblock \bibinfo{title}{{\"Uber} die magnetischen {Momente der Atomkerne}},
\newblock \bibinfo{journal}{Z. Physik} \bibinfo{volume}{106}
  (\bibinfo{year}{1937}) \bibinfo{pages}{358--361}.
%Type = Article
\bibitem[{Bender et~al.(2002)Bender, Dobaczewski, Engel, and
  Nazarewicz}]{(Ben02d)}
\bibinfo{author}{M.~Bender}, \bibinfo{author}{J.~Dobaczewski},
  \bibinfo{author}{J.~Engel}, \bibinfo{author}{W.~Nazarewicz},
\newblock \bibinfo{title}{{Gamow-Teller} strength and the spin-isospin coupling
  constants of the {Skyrme} energy functional},
\newblock \bibinfo{journal}{Phys. Rev. C} \bibinfo{volume}{65}
  (\bibinfo{year}{2002}) \bibinfo{pages}{054322}.
%Type = Article
\bibitem[{Sassarini et~al.(2021)Sassarini, Dobaczewski, Bonnard, and {Garcia
  Ruiz}}]{(Sas21)}
\bibinfo{author}{P.~L. Sassarini}, \bibinfo{author}{J.~Dobaczewski},
  \bibinfo{author}{J.~Bonnard}, \bibinfo{author}{R.~F. {Garcia Ruiz}},
\newblock \bibinfo{journal}{arXiv:2111.04675}  (\bibinfo{year}{2021}).

\end{thebibliography}
\end{document}